\documentclass[longauth]{aa}  %
\usepackage{graphicx}
\usepackage{txfonts}
\usepackage{hyperref}
\usepackage{amsmath}
\usepackage[separate-uncertainty=true, multi-part-units=single]{siunitx}
\usepackage{comment}
\usepackage{amstext}

\def\ltsima{$\; \buildrel < \over \sim \;$}
\def\simlt{\lower.5ex\hbox{\ltsima}}
\def\gtsima{$\; \buildrel > \over \sim \;$}
\def\simgt{\lower.5ex\hbox{\gtsima}}
\def\gsimeq
{\hbox{\raise0.5ex\hbox{$>\lower1.06ex\hbox{$\kern-1.07em{\sim}$}$}}}
\def\lsimeq
{\hbox{\raise0.5ex\hbox{$<\lower1.06ex\hbox{$\kern-1.07em{\sim}$}$}}}

\def\xmm{{\it XMM-Newton }}

\def\xmm{{\it XMM-Newton}}
\def\chandra{{\it Chandra}}

\def\fermi{{\it Fermi}}

\def\nustar{{\it NuSTAR}}
\def\gravity{{GRAVITY}}
\def\spitzer{{\it Spitzer}}
\def\sinfoni{{\it SINFONI}}

\def\aap{A\&A}

\def\apjl{ApJ}
\def\apjs{ApJS}

\def\sgr{Sgr~A$^{\star}$}

\def\axj{AX~J1745.6-2901}

\def\plcg{PLCool$\gamma_{max}$}

\def\sgras{Sgr~A$^\star$}

\begin{document}

   \title{Constraining particle acceleration in \sgras\ with simultaneous \gravity, \spitzer, \nustar\ and \chandra\ observations} 
   \titlerunning{Multiwavelength observations of \sgras}

   \author{GRAVITY Collaboration\thanks{GRAVITY is developed
    in a collaboration by the Max Planck Institute for
    extraterrestrial Physics, LESIA of Observatoire de Paris/Universit\'e PSL/CNRS/Sorbonne Universit\'e/Universit\'e de Paris and IPAG of Universit\'e Grenoble Alpes /
    CNRS, the Max Planck Institute for Astronomy, the University of
    Cologne, the CENTRA - Centro de Astrofisica e Gravita\c c\~ao, and
    the European Southern Observatory. $\,\,\,\,\,\,$ Corresponding authors: S. D. von Fellenberg (email sefe@mpe.mpg.de), G. Ponti (email ponti@mpe.mpg.de), Y. Dallilar (email ydalillar@mpe.mpg.de) and G. Witzel (email gwitzel@mpifr-bonn.mpg.de )
    }:
   R.~Abuter\inst{8}
\and A.~Amorim\inst{6,12}
\and M.~Baub\"ock\inst{1}
\and F.~Baganoff\inst{20}
\and J.P.~Berger\inst{5,8}
\and H.~Boyce\inst{21, 22}
\and H.~Bonnet\inst{8}
\and W.~Brandner\inst{3}
\and Y.~Cl\'{e}net\inst{2}
\and R.~Davies\inst{1}
\and P.T.~de~Zeeuw\inst{10,1}
\and J.~Dexter\inst{14,1}
\and Y.~Dallilar\inst{1}
\and A.~Drescher\inst{1,17}
\and A.~Eckart\inst{4,18}
\and F.~Eisenhauer\inst{1}
\and G.G.~Fazio\inst{19} 
\and N.M.~F\"orster~Schreiber\inst{1} 
\and K. Foster\inst{33}
\and C.~Gammie\inst{25}
\and P.~Garcia\inst{7,12}
\and F.~Gao\inst{1}
\and E.~Gendron\inst{2}
\and R.~Genzel\inst{1,11}
\and G.~Ghisellini\inst{13} 
\and S.~Gillessen\inst{1}
\and M.A.~Gurwell \inst{19}
\and M.~Habibi\inst{1}
\and D.~Haggard\inst{21, 22} 
\and C.~Hailey\inst{29}
\and F. A. Harrison\inst{33}
\and X.~Haubois\inst{9}
\and G.~Heissel\inst{2} 
\and T.~Henning\inst{3}
\and S.~Hippler\inst{3}
\and J.L.~Hora\inst{19} 
\and M.~Horrobin\inst{4}
\and A.~Jim\'enez-Rosales\inst{1}
\and L.~Jochum\inst{9}
\and L.~Jocou\inst{5}
\and A.~Kaufer\inst{9}
\and P.~Kervella\inst{2}
\and S.~Lacour\inst{2}
\and V.~Lapeyr\`ere\inst{2}
\and J.-B.~Le~Bouquin\inst{5}
\and P.~L\'ena\inst{2}
\and P.J.~Lowrance\inst{23} 
\and D.~Lutz\inst{1}
\and S.~Markoff\inst{28}
\and K.~Mori\inst{29}
\and M.R.~Morris\inst{24}
\and J.~Neilsen\inst{32}
\and M.~Nowak\inst{16,2}
\and T.~Ott\inst{1}
\and T.~Paumard\inst{2}
\and K.~Perraut\inst{5}
\and G.~Perrin\inst{2}
\and G.~Ponti\inst{14,1}
\and O.~Pfuhl\inst{8,1}
\and S.~Rabien\inst{1}
\and G.~Rodr\'iguez-Coira\inst{2}
\and J.~Shangguan\inst{1}
\and T.~Shimizu\inst{1}
\and S.~Scheithauer\inst{3}
\and H.A.~Smith\inst{19} 
\and J.~Stadler\inst{1}
\and D.~K.~Stern\inst{34} 
\and O.~Straub\inst{1}
\and C.~Straubmeier\inst{4}
\and E.~Sturm\inst{1}
\and L.J.~Tacconi\inst{1}
\and F.~Vincent\inst{2}
\and S.~von~Fellenberg\inst{1}
\and I.~Waisberg\inst{15,1}
\and F.~Widmann\inst{1}
\and E.~Wieprecht\inst{1}
\and E.~Wiezorrek\inst{1} 
\and S.P.~Willner\inst{19} 
\and G.~Witzel\inst{18} 
\and J.~Woillez\inst{8}
\and S.~Yazici\inst{1,4}
\and A.~Young\inst{1}
\and S.~Zhang\inst{30} 
\and G.~Zins\inst{9}
}
   \authorrunning{\gravity\ collaboration et al.}
\institute{
Max Planck Institute for extraterrestrial Physics,
Giessenbachstrasse~1, 85748 Garching, Germany
\and LESIA, Observatoire de Paris, Universit\'e PSL, CNRS, Sorbonne Universit\'e, Universit\'e de Paris, 5 place Jules Janssen, 92195 Meudon, France
\and Max Planck Institute for Astronomy, K\"onigstuhl 17, 
69117 Heidelberg, Germany
\and $1^{\rm st}$ Institute of Physics, University of Cologne,
Z\"ulpicher Strasse 77, 50937 Cologne, Germany
\and Univ. Grenoble Alpes, CNRS, IPAG, 38000 Grenoble, France
\and Universidade de Lisboa - Faculdade de Ci\^encias, Campo Grande,
1749-016 Lisboa, Portugal 
\and Faculdade de Engenharia, Universidade do Porto, rua Dr. Roberto
Frias, 4200-465 Porto, Portugal 
\and European Southern Observatory, Karl-Schwarzschild-Strasse 2, 85748
Garching, Germany
\and European Southern Observatory, Casilla 19001, Santiago 19, Chile
\and Sterrewacht Leiden, Leiden University, Postbus 9513, 2300 RA
Leiden, The Netherlands
\and Departments of Physics and Astronomy, Le Conte Hall, University
of California, Berkeley, CA 94720, USA
\and CENTRA - Centro de Astrof\'{\i}sica e
Gravita\c c\~ao, IST, Universidade de Lisboa, 1049-001 Lisboa,
Portugal
\and INAF-Osservatorio Astronomico di Brera, Via E. Bianchi 46, I-23807 Merate (LC), Italy
\and Department of Astrophysical \& Planetary Sciences, JILA, Duane Physics Bldg., 2000 Colorado Ave, University of Colorado, Boulder, CO 80309, USA
\and Department of Particle Physics \& Astrophysics, Weizmann Institute of Science, Rehovot 76100, Israel
\and Institute of Astronomy, Madingley Road, Cambridge CB3 0HA, UK
\and Department of Physics, Technical University Munich, James-Franck-Strasse 1,  85748 Garching, Germany
\and Max Planck Institute for Radio Astronomy, Auf dem H\"ugel 69, 53121 Bonn, Germany
\and Center for Astrophysics \textbar\ Harvard \& Smithsonian,
 60 Garden St., Cambridge, MA 02138 USA
\and MIT Kavli Institute for Astrophysics and Space Research,
  Cambridge, MA 02139, USA
\and Department of Physics, McGill University, 3600
University St., Montreal, QC H3A 2T8, Canada
\and McGill Space Institute, 3550 University St., Montreal, QC H3A 2A7, Canada
\and Spitzer Science Center, California Institute of Technology, Pasadena, CA 91125 USA
\and UCLA Galactic Center Group, Physics and Astronomy Department, University of California, Los Angeles, CA 90024
\and Department of Astronomy, University of Illinois, 1002 West Green Street, Urbana, IL 61801, USA
\and Department of Astronomy, Boston University, Boston, MA 02215, USA
\and Steward Observatory, University of Arizona, 933 North Cherry Avenue, Tucson, AZ 85721, USA
\and Anton Pannekoek Institute for Astronomy, University of Amsterdam, 1098 XH Amsterdam, The Netherlands
\and Columbia Astrophysics Laboratory, Columbia University, 550 West 120th Street, Room 1027, New York, New York 10027, USA
\and Bard College Physics Program, 30 Campus Road, Annandale-On-Hudson, NY 12504, USA
\and SOFIA Science Center, Moffett Field, CA USA
\and Villanova University, Department of Physics, Villanova, PA 19085, USA
\and Cahill Center for Astronomy and Astrophysics, California
Institute of Technology, Pasadena, CA 91125, USA
\and Jet Propulsion Laboratory, California Institute of Technology, 4800 Oak Grove Drive, MS 169-224, Pasadena, CA
91109, USA
}

   \date{Received \today; accepted future}

 
  \abstract
    {We report the time-resolved spectral analysis of a bright near-infrared and moderate X-ray flare of \sgras. 
    We obtained light curves in the $M$-, $K$-, and $H$-bands in the mid- and near-infrared and in the $2-8~\mathrm{keV}$ and $2-70~\mathrm{keV}$ bands in the X-ray. The observed spectral slope in the near-infrared band is $\nu L_\nu\propto \nu^{0.5\pm0.2}$; the spectral slope observed in the X-ray band is $\nu L_\nu \propto \nu^{-0.7\pm0.5}$. 
    Using a fast numerical implementation of a synchrotron sphere with constant radius, magnetic field and electron density (i.e.\ a one-zone model), we tested various synchrotron and synchrotron self-Compton scenarios. The observed near-infrared brightness and X-ray faintness, together with the observed spectral slopes, pose challenges for all models explored. We rule out a scenario in which the near-infrared emission is synchrotron emission and the X-ray emission is synchrotron self-Compton. Two realizations of the one-zone model can explain the observed flare and its temporal correlation: one-zone model in which both the near-infrared and X-ray luminosity are produced by synchrotron self-Compton and a model in which the luminosity stems from a cooled synchrotron spectrum. Both models can describe the mean SED and temporal evolution similarly well. In order to describe the mean SED, both models require specific values of the maximum Lorentz factor $\gamma_{max}$, which however differ by roughly two orders of magnitude: the synchrotron self-Compton model suggests that electrons are accelerated to $\gamma_{max}\sim 500$, while cooled synchrotron model requires acceleration up to $\gamma_{max}\sim5\times 10^{4}$. The synchrotron self-Compton scenario requires electron densities of $10^{10}~\mathrm{cm^{-3}}$ much larger than typical ambient densities in the accretion flow. Furthermore, it requires a variation of the particle density inconsistent with average mass-flow rate inferred from polarization measurements, and can therefore only be realized in an extraordinary accretion event. In contrast, assuming a source size of $1R_s$, the cooled synchrotron scenario can be realized with densities and magnetic fields comparable with the ambient accretion flow. For both models, the temporal evolution is regulated through the maximum acceleration factor $\gamma_{max}$, implying that sustained particle acceleration is required to explain at least a part of the temporal evolution of the flare.}


   \maketitle
%

\section{Introduction} 

It is believed that most galaxies harbour at least one 
supermassive black hole (BH) at their centre (Kormendy \& Ho 2013). 
However, only a small fraction are accreting at a high rate and appear as active galactic nuclei. The vast majority are quiescent and therefore inaccessible to us. 
One exception is \sgras. 
Located only 8.27~kpc from us \citep{GRAVITYCollaboration2019, Do2019, GravityCollaboration2021}, \sgras\ is the closest supermassive BH, with a mass of $(4.297\pm0.013)~\mathrm{M_\odot}$ and a corresponding Schwarzschild radius of $R_S=2GM_{BH}/c^2{\sim} 1.3\times10^{10}~\mathrm{m}$. 
Because it is so close, \sgras\ appears orders of magnitudes brighter than any other supermassive BH in quiescence despite its faint X-ray flux of ${\sim}2\times10^{33}$~erg~s$^{-1}$ \citep{Baganoff2003}. 
Therefore, \sgras\ offers a unique opportunity to study the physics of accretion in quiescent systems. 

The majority of \sgras's steady radiation is emitted at sub-mm frequencies, most likely produced by optically thick synchrotron emission originating from relativistic thermal electrons in the central ${\sim}10$ Schwarzschild radii ($R_S$) at temperatures of $T_e{\sim}$ a few $10^{11}~K$ and densities  $n_e{\sim}10^{7}~\mathrm{cm^{-3}}$, 
embedded in a magnetic field with strength of $\sim10-50$~G \citep{LoebWaxman2007, VonFellenberg2018, Bower2019}. This implies that the accretion flow at a few Schwarzschild radii from the black hole is strongly magnetised. Indeed, for an ambient magnetic field strength of $B{\sim}40$~G and ambient $n_e\sim10^6~\mathrm{cm^{-3}}$, we estimate a plasma parameter $\beta$ of ${\sim}0.04$ (comparing the thermal pressure of the gas with the magnetic pressure), and $\sigma_{th}{\sim}15$ (comparing the magnetic field energy with the thermal energy).

In the X-ray band, \sgras\ appears as a faint ($L_{2-10~keV}\sim2\times10^{33}$~erg~s$^{-1}$) extended source with a size, ${\sim}1^{\prime\prime}$, comparable to the Bondi radius, emitting via bremsstrahlung emission from a hot plasma with $T_e\sim7\times10^7$~K and $n_e\sim100~\mathrm{cm^{-3}}$ \citep{Quataert2002, Baganoff2003, Xu2006}. In the X-ray band, \sgras\ occasionally shows sudden rises (flares) of up to 1-2 orders of magnitudes, suggesting individual and distinct events, randomly punctuating an otherwise quiescent source \citep{Baganoff2001, Porquet2003, Porquet2008, Neilsen2013, Ponti2015, Bouffard2019}.
X-ray flares are associated with bright flux excursions in the near-infrared (IR) band, which also led to the definition of the latter as flares \citep{Genzel2003, Ghez2004}.
However, the IR emission is continuously varying \citep{Do2009, Meyer2009, Witzel2018}.

In 2018, \cite{GravityCollaboration2018_orbital} reported the first detection of an orbital signature in the centroid motion of three \sgras{} flares. The centroid motion of the three flares is consistent with a source on a relativistic orbit around the black hole. Using a fully general relativistic model of a ``hot spot'', the authors derived a typical orbital radius of around ${\sim} 4.5~\mathrm{R_s}$, and constrained the emission regions to ${\sim} 2.5~\mathrm{R_s}$, and a viewing angle of $i{\sim} 140\deg$ (the inclination of the orbital plane to the line of sight). This model was extended by \cite{GravityCollaboration2020_orbital}, who showed that the flare light curves may be modulated by Doppler boosting on the order a few tens of percent. The polarimetric analysis of these flares showed consistent results \citep{GravityCollaboration2020_polariflares}. These findings further cement the picture of flares originating from localized regions of the accretion flow in which particles are heated or accelerated.

However, the radiative mechanism powering flares is still disputed. The most common proposed mechanisms are: synchrotron with a cooling break; synchrotron self-compton (SSC); inverse compton (IC); and Synchrotron \citep{Markoff2001, Yuan2003, Eckart2004, Eckart2006, Eckart2008, Eckart2009, Eckart2012, YusefZadeh2006,YusefZadeh2008,YusefZadeh2009, Hornstein2007, Marrone2008, Dodds-Eden2009, Dodds-Eden2010, Trap2011, Dibi2014, Barriere2014}.
Simultaneous determination during an X-ray flare of the photon index ($\Gamma$) in the near-infrared (NIR; $\Gamma_{\rm{IR}}$) and X-ray ($\Gamma_{\rm{X}}$) bands allows us to discriminate synchrotron and synchrotron with a cooling break from the other radiative mechanisms. 
It is expected that $\Gamma_{\rm{X}}=\Gamma_{\rm{IR}}$ or $\Gamma_{\rm{X}}=\Gamma_{\rm{IR}}+0.5$ for the synchrotron and synchrotron with a cooling break model, respectively \citep{Kardashev1962, Pacholczyk1970, Dodds-Eden2010, Ponti2017}. Any other value would favour either SSC or IC scenarios. 

Thanks to an extensive multi-wavelength monitoring campaign covering from IR (with \sinfoni) to X-ray (with \xmm+\nustar), \cite{Ponti2017} observed a very bright NIR and X-ray flare in August 2014. 
The radiative mechanism was consistent with synchrotron emission all the way from IR to X-ray, therefore implying the presence of a powerful accelerator (with $\gamma_{max}>10^{5-6}$) and an evolving cooling break and high energy cutoff in the distribution of accelerated particles. 
This demonstrated that, at least for that flare, synchrotron emission with a cooling break and a varying high energy cutoff is a viable mechanism. 

To obtain a better insight into \sgras's flaring activity, we deployed a large multi-wavelength campaign in July 2019. 
The campaign was built around a core of three strictly simultaneous 16~hr \chandra\ and \spitzer\ observations covering \sgras's emission in the soft X-ray and $M$-band (PI G.G.~Fazio). 
In addition two long \nustar\ exposures were performed to simultaneously cover the entire campaign in the hard X-ray band. 
Finally, a ${\sim}6.5$~hr observation with the VLTI-\gravity\ interferometer was performed in the night between July $17^{th}$ and $18^{th}$, expanding the campaign to the $K$ and $H$-bands. For simplicity, we refer to the IR observations by the observing band most similar with the effective wavelength of the observations throughout the paper. Table \ref{data} reports the effective wavelength.
Observations with the Submillimeter Array (Witzel et al. 2021) were approved but not executed owing to a number of factors including weather and limited access to the array during the summer of 2019.
During the time window when all instruments were active, we caught a bright infrared and moderate X-ray flare. We report here the characterisation and evolution of the IR to X-ray spectral energy distribution during the flare and the implications for our understanding of particle acceleration during \sgras's flares. 

\section{Data reduction} 
\label{datared}

\subsection{Basic assumptions}

Throughout this paper we assume a distance to \sgras\ of 8.249 kpc and
a mass $M_{\rm{BH}} = 4.26\times 10^{6} M_{\odot}$ (Gravity Collaboration 2020). 
Errors and upper limits quoted are at the $1\sigma$ and 90\% confidence 
level, respectively. The X-ray data were initially fitted with {\sc xspec} v. 12.10.1f, employing the Cash statistics in spectral fits (Cash 1979). 
Throughout our analysis and discussion we make the following assumptions:

\begin{itemize}
    \item Effects of beaming are negligible.
    \item Emission is dominated by a single emitting zone.
    \item Unless otherwise stated, we follow \cite{Do2009} and assume a constant escape time of the synchrotron emitting electrons equal to $t_{esc} = 120$ s.
\end{itemize}

\begin{table*}
\centering
\begin{tabular}{ l c c c c c c}
\hline
\hline
Instrument & OBSID & Start   & Start & Exp  &Energy & Wavelength\\
                  &             & (UTC)  & (MJD) & (ks)  &  & \\
\hline
\chandra\ & 22230 & 2019-07-17 22:51:26  & 58681.9524& 57.6 & 2--8keV &  6.2--1.6\SI{}{\angstrom}\\
                & 20446 & 2019-07-21 00:00:14  & 58685.0002 & 57.6  & 2--8keV & 6.2--1.6\SI{}{\angstrom}\\
                & 20447 & 2019-07-26 01:32:40  & 58690.0639& 57.6  & 2--8keV & 6.2--1.6\SI{}{\angstrom}\\
\hline
\nustar\ & 30502006002 & 2019-07-17 21:51:09  & 58681.9105& 38.6  & 2--70keV& 6.2--0.2\SI{}{\angstrom}\\
             & 30502006004 & 2019-07-26 00:41:09  & 58690.0286 & 34.8  & 2--70keV & 6.2--0.2\SI{}{\angstrom}\\
\hline
\gravity\ & 0103.B-0032(D) & 2019-07-17 23:32:55  &58681.9812 & 21.6 & 0.7--0.8eV & 2.2--1.65 $\mu m$ \\
\hline
\spitzer\ & 69965312 & 2019-07-17 23:21:33 & 58681.9733 & 17.6  & 0.3eV& 4.5 $\mu m$ \\ 
          & 69965568 & 2019-07-18 07:25:02 & 58682.3091 & 17.6  & 0.3eV& 4.5 $\mu m$ \\ 
\hline
\end{tabular} 
\caption{Datasets analysed in this work. 
The table reports the instrument used, the identification number of the dataset, the start time of the observation, the total exposure and the energy bands and effective wavelengths of the different instruments}.
\label{data}
\end{table*} 

\subsection{\textbf{\textit{Chandra}}}
\label{sec:chandra}
A series of three \chandra\ \citep{Weisskopf2000} observations has been analysed (see Tab. \ref{data}). To enhance sensitivity and reduce the effects of pile-up during flares of \sgras, the observations were taken with ACIS-S at the focus \citep{Garmire2003}. Only one CCD was active (S3) with 1/8 subarray (i.e. 128 rows) and no grating applied.
The data have been reduced with standard tools from the {\sc ciao} analysis suite, version 4.12 \citep{Fruscione2006} and calibration database v4.9.3, released on October 16$^{th}$ 2020.
The data from each observation were reprocessed applying the {\sc chandra\_repro} script with standard settings. Barycentric corrections with the task {\sc axbary} were applied to the events files, the aspect solution, and all products. 
To match the exposure of the \gravity\ light curves, we computed light curves in the 2--8~keV, 2--4~keV, and 4--8~keV bands with 380~s time bins, following the \gravity\ exposure time of 320~s plus a dead time of approximately 60~s. 
Considering the small number of events during quiescence, we display the count rates following the Gehrels approximation ($\sqrt{(N+0.75)}+1$; \citealt{Gehrels1986}). 

During OBSID 22230, we observed a peak count rate of 0.09 ph~s$^{-1}$ in the 2--8~keV band. 
Given the instrumental set up, pile-up effects are negligible even at the peak (e.g.\ \citealt{Ponti2015}).  
By using the \cite{Ponti2015} conversion factors, we estimate a total observed (absorbed) energy of ${\sim}3.2\times10^9$ erg released during the flare in the 2--8~keV band. Following the classification of \cite{Ponti2015}, this flare belongs to the group of moderate flares in the X-ray band. 

Photons from \sgras\ were extracted from a circular region of $1.25''$ radius. 
The spectrum of the flare was extracted with {\sc specextract} within the time interval {\sc mjd = 58682.134:58682.148} (see dotted lines in \autoref{lcflare}) and contains a total of 72 photons in the 2-10~keV band. 
The background spectrum was extracted from the same source region but from the events file accumulated during {\sc obsid} 20447, during which no flare of \sgras\ was detected. 

\subsection{\textbf{\textit{NuSTAR}}}

To study the flare characteristics in the hard X-ray band, we analysed the two NuSTAR \citep{Harrison2013} observations taken in July 2019 in coordination with \gravity, \chandra, and \spitzer\ (Table \ref{data}). 
We processed the data using the \nustar\ Data Analysis Software {\it NUSTARDAS} and HEASOFT v. 6.28, and CALDB v20200912, filtered for periods of high instrumental background due to South Atlantic anomaly passages and known bad detector pixels. The data were barycenter corrected. 
Products were extracted from a region of radius $20''$ centered on the position of \sgras\ using the tool {\sc nuproducts} within the intervals shown in Fig. \ref{lcflare}. 
The background spectra were extracted from the same region in the off-flare intervals within the same observation. 
In particular, the background spectrum has been integrated for each orbit during which no X-ray flares nor bright IR flux excursions have been observed in the \nustar+\chandra\ and \gravity+\spitzer\ light curves (Boyce et al. in prep.), resulting in a net exposure time of ${\sim}30$~ks. 
Because part of the FPMB instrument is affected by stray light due to a Galactic center X-ray transient outside of the field of view, we present the analysis of the FPMA data only. 
The results from FPMB are consistent with the ones presented here.
The light curves were accumulated in the 3--10 keV band and with 380~s time bins for comparison with the \gravity\ data. 
Bins with small fractional exposures have been removed. 

\subsection{\textbf{\textit{Spitzer/IRAC}}\label{sec:spitzeer data }}
The observations were obtained using the IRAC instrument \citep{Fazio2004} on the \spitzer{} Space Telescope \citep{Werner2004}. The observations were part of the \spitzer\ program 14026 \citep{Fazio2018}, which observed \sgr{}\ at $4.5~\mathrm{\mu m}$ during three epochs of ${\sim}16$ hours each in 2019 July. The observing sequence included an initial mapping operation and then two successive 8-hour staring-mode observations, each using the ``PCRS peak-up'' to center \sgr{} on pixel (16,16) of the subarray. We used a similar data pipeline as described by \cite{Hora2014}, \cite{Witzel2018}, and \cite{Boyce2019} to derive differential flux measurements. Modifications to the procedure for reduction and calibration of the light curves were necessary because of the larger pointing drift compared to previous observations (about one full pixel over the first three hours of the staring observation). The procedure was modified to transition to the neighboring pixel for the flux measurement when the drift moved \sgr{} into that pixel, roughly one hour after the start of the stare. Also because of the large drift, we derived a new calibration curve that would be valid over the larger range. We used observations of standard stars previously obtained for the subarray ``sweet spot'' calibration \citep{Ingalls2012}, and found that a fifth-degree polynomial using the distance from the center of the pixel and central pixel flux density provided an acceptable fit to the total flux density of a point source with a standard deviation consistent with the S/N of the observations. 

The uncertainty of the \spitzer\ light curve was estimated by computing the standard deviation of the light curve sections where the \gravity\ $K$-band flux was low. Because the light curve shows residual artifacts from the imperfect background subtraction, we scaled the standard deviation such that low-flux parts of the light curve have $\chi^2_{red}=1$ with respect to zero mean flux. The flux was de-reddened using the \cite{Fritz2011} extinction values reported in Table \ref{tab:extinction}. Because the \spitzer\ light curve was derived through differential photometry, we need to add a flux offset. We used the method described by \cite{Witzel2018} to account for the flux offset but used the median $K$-band flux derived by \cite{GRAVITYCollaboration2020flux}. Explicitly, we added $1.8\pm0.3~\mathrm{mJy}$ to all differential flux measurements of \spitzer.

\subsection{\textbf{\textit{GRAVITY}}\label{sec:gravity data}}
The interferometric $K$-band flux density was determined in the same way as by \cite{GRAVITYCollaboration2020flux}. The values reported are the coherent flux values corrected for the contribution of the star S2. We neglected the contribution of the star S62, which amounts to a constant flux of ${\sim}0.1~\mathrm{mJy}$. For the details of the flux determination, see \cite{GRAVITYCollaboration2020flux}.

The $H$-band flux was determined from aperture photometry of the deconvolved acquisition camera images. The acquisition camera of GRAVITY is normally used for the acquisition of the observation as well as the field and pupil tracking for each of the four unit telescopes. In order to use the aquistion camera images for science, we averaged the four images\footnote{The aquistion camera pipeline will be made available under \url{https://github.com/Sebastiano-von-Fellenberg/AquisitionCamera}. It has been written by SvF and Giuila Folchi.}. The images were bad-pixel-corrected and dark-subtracted. We approximated the PSF of the images by a Gaussian. The parameters of the Gaussian were determined by fitting a Gaussian model to the bright star S10 and we used this PSF model to deconvolve the images using the Lucy-Richardson algorithm implemented in dpuser\footnote{https://www.mpe.mpg.de/~ott/dpuser/}. 

In both $K$- and $H$-band we measured the flux ratio of Sgr~A* relative to S2. Because Sgr~A* is a much redder source than S2  \citep{Genzel2010}, we have to take the difference in spectral index into account. For the $K$-band this was achieved by fitting a power-law spectrum to both sources and determining the flux at $2.2~ \mathrm{\mu m}$. 
For the $H$-band, we accounted for this difference in spectral index by assuming that the reddened flux from both sources is described by a power law. We used NACO photometry of S2 to determine the reddened spectral slope of S2. By using the observed flux ratio in the $H$- and $K$-bands and the transmission curve of the acquisition camera detector, we derived the effective wavelength of Sgr~A* in the $H$-band: $\lambda_{\rm{Sgr~A*}}\sim 1.63~\mathrm{\mu m}$. Once the effective wavelength was determined, we used the observed flux ratio in $H$- and $K$-band to determine the flux density of Sgr~A* in the $H$-band. The details of this are outlined in \autoref{appendix:hband flux}.

\subsection{Extinction\label{sec:extinction}}
The Galactic Center is a highly extincted region, with an approximately broken-power-law extinction $A(\lambda)$ between $1.2~\mathrm{\mu m}$ and $8~\mathrm{\mu m}$ \citep{Fritz2011}. The extinction is a major source of uncertainty for our analysis because even a small variation in the power-law extinction slope leads to a large change in our measured IR spectral slope. The hydrogen column density is similarly a key ingredient in the derivation of the X-ray absorption and thus the modeling of the X-ray spectral slope. Moreover, the hydrogen column density and the IR extinction are related but independently determined. This may therefore lead to a systematic offset between NIR and X-ray observations. 
\subsubsection{IR extinction\label{sec:nir extinction}}
We used the extinction model from \cite{Fritz2011}, who used the hydrogen emission lines observed with SINFONI at the VLT to derive a broken-power-law extinction curve. This allows us to drop the uncertainty on the absolute calibration and only propagate the uncertainty on the power law exponents. The authors also provided extinction values for NACO and \spitzer{}, tabulated in \autoref{tab:extinction}. We neglected the uncertainty due to the difference in filter response between NACO and the two GRAVITY bands. 

\begin{table}[]
    \centering
    \begin{tabular}{c|c|}
         Band & \cite{Fritz2011} \\
         \hline
         $\mathrm{A_H}$ &$4.21 \pm 0.08$\\
         $\mathrm{A_{Ks}}$&$2.42 \pm  0.002$\\
         $\mathrm{A_M}$ &$0.97 \pm 0.03$
    \end{tabular}
    \caption{Extinction values of \citealt{Fritz2011} in magnitudes. The uncertainties of \citealt{Fritz2011} have been propagated only taking into account only the uncertainty of the spectral slope.}
    \label{tab:extinction}
\end{table}


\subsubsection{X-ray extinction}

The observed X-ray spectrum is distorted by the combination of 
absorption and dust scattering. 
The latter effect produces a halo of emission, which is typically 
partially included within the limited extraction region used to compute the 
spectrum of \sgras. We fitted the dust's scattering halo with the model {\sc fgcdust} in {\sc XSpec} \citep{Jin2017, Jin2018}, and it was assumed to be the same as the ‘foreground’ component along 
the line of sight towards \axj\ \citep{Jin2017, Jin2018}. 

We fit the absorption affecting the X-ray spectra  with the model 
{\sc tbabs} (see \citealt{WilmsAllenMcCray2000}) with the cross-sections of \cite{Verner1996} and abundances from \cite{Wilms2000}. 
Figure \ref{DeAb} shows the impact of 
the different assumptions for the column density on the X-ray spectral slope. 
As \cite{Ponti2017}, we assumed a column density of $N_{\rm{H}}=1.6\times10^{23}$ cm$^{-2}$.

\section{Light curves}
\begin{figure*}
\includegraphics[ width=0.95\textwidth,angle=-0]{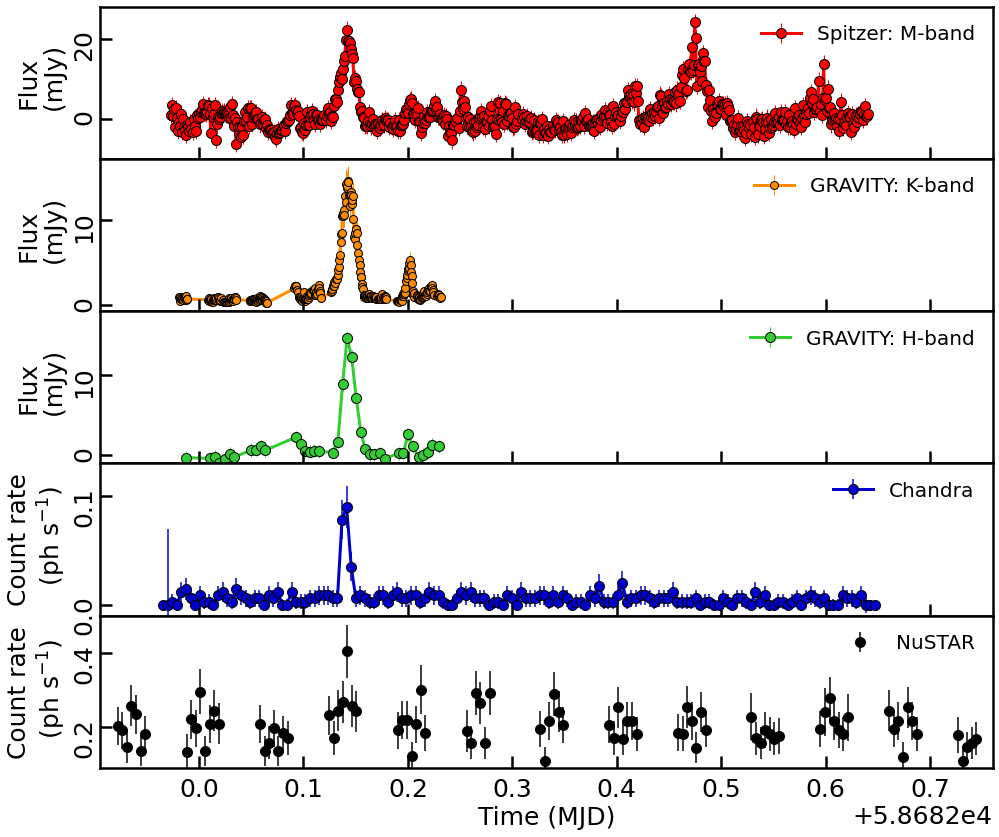}
\caption{X-ray and IR light curves of the multi-wavelength observations performed on 2019 July $18^{\rm th}$. The \spitzer\ (red), \gravity\ $K$ (orange) and $H$-band (green), \chandra\ (blue), and \nustar (black) data. The \spitzer\ light curves show the differential flux density. The NIR flux densities have been corrected for extinction using the values in Table \ref{tab:extinction}.}
\label{lc}
\end{figure*}
Fig. \ref{lc} shows the full duration of the multi-wavelength campaign performed on 2019 July 17$^{\rm th}$--18$^{\rm th}$. The \spitzer\ and \gravity\ light curves follow each other very well. The \spitzer\ light curve shows IR flares in excess of 5~mJy. 
In particular, two $F_M\gsimeq15~\mathrm{mJy}$ and $t\gsimeq30~\mathrm{min}$ IR flares are observed by \spitzer\ at $MJD{\sim}58682.14$ and ${\sim}58682.47$. 
However, only the first IR flare has a detectable X-ray counterpart (Fig. \ref{lc}). 
Which suggests that one or more additional parameters are required to control the "X-ray loudness" of the IR flares. 

\begin{figure*}
\centering
\includegraphics[width=0.95\textwidth,angle=-0]{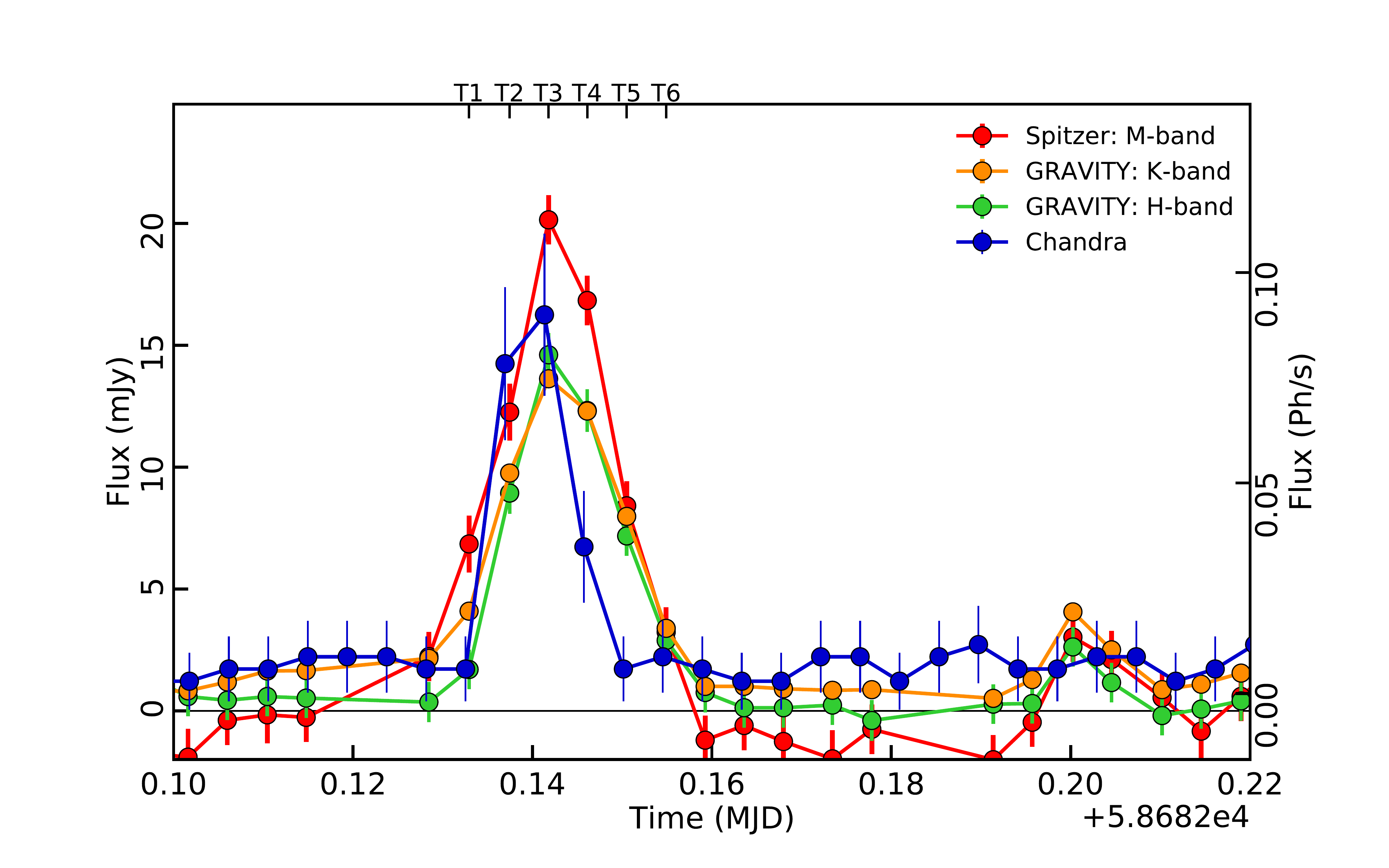}
\caption{X-ray and IR light curves of the flare detected on 2019 July $18^{th}$. 
The blue points show the \chandra\ light curve in the 2--8~keV band. The red, orange and green points show \spitzer\ ($M$-band), the \gravity\ $K$-band and $H$-band light curves corrected for extinction, respectively. The bold ticks on the top abscissa labeled T1, T2, T3, T4, T5, and T6 mark the times that will be used in the subsequent analysis.}
\label{lcflare}
\end{figure*}

\autoref{lcflare} shows a zoom-in of the light curves of the bright IR flare with X-ray counterpart detected on July 18$^{\rm th}$ 2019. 
As discussed by Boyce et al.\ (in prep.), the flare occurred nearly simultaneously in the two bands, with the X-ray peak occurring at the maximum of the IR emission. 
The X-ray flare, as observed by \chandra, was shorter (${\sim}19$~min duration) than its IR counterpart (${\sim}38$~min duration). 
A shorter duration of the X-ray flare has been observed before (e.g. \citealt{Dodds-Eden2010}; \citealt{Dodds-Eden2011}). 

At the start of the flare (T1\footnote{T1 stands for the first time interval of the time resolved analysis. } ${\sim} 58682.133$) emission was observed in the $K$- and $M$-bands (${\sim}5$~mJy) with simultaneous $H$-band emission but no excess above quiescence in the X-ray band. 
Soon after, the X-ray band rose very rapidly (T2). It then decayed quickly back to quiescence, while the IR flux rose and decayed more gently (Fig. \ref{lcflare}). Indeed, when X-ray emission reached quiescence, the IR flux density was still above ${\sim}8$~mJy in every IR band (Fig. \ref{lcflare}; T5 and T6). 


\section{The multi-wavelength flare in context}
\begin{figure}
    \centering
    \includegraphics[width=0.45\textwidth,angle=-0]{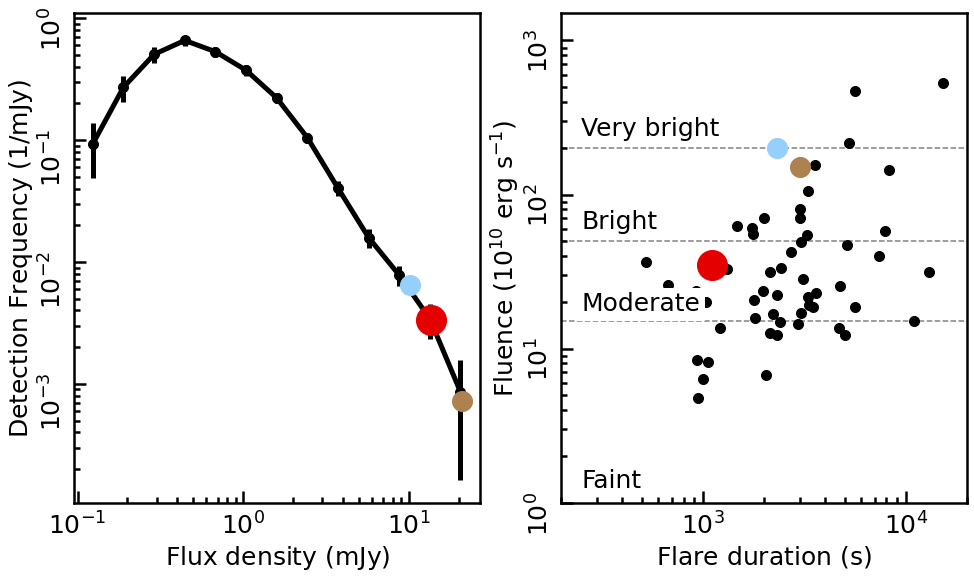}
    \caption{Left: The GRAVITY $K$-band flux density distribution as reported in \cite{GRAVITYCollaboration2020flux} and the peak flux densities of three bright flares. The red point indicates the peak flux density of the flare analyzed in this paper. The light blue point indicates the peak flux reported by \cite{Ponti2017} observed with SINFONI. The light brown point is the peak L'-band flux density scaled to $2.2~\mathrm{\mu m}$ assuming a flux density scale $F_{K band} = F_{L' band}\cdot(\nu_K/\nu_{L'})^{-0.5}$.
Right: Duration and fluence of all flares of \sgras\ detected by \xmm\ and \chandra\ before 2015 (see \citealt{Neilsen2013, Ponti2015}). Partial (i.e.\ only partially covered) and dubious flares have been omitted. As in the left plot, the red, light blue, and dark blue circles show the duration and fluence of the X-ray flares investigated here, by \cite{Ponti2017} and by \cite{Dodds-Eden2009}. }
    \label{fig:fluence}
\end{figure}

The IR flare reported in this paper is among the brightest ever observed. It is the third brightest flare observed with GRAVITY, although it is significantly shorter than the flares observed in 2019. The left panel of \autoref{fig:fluence} shows the flux distribution of \sgras{} \citep{GRAVITYCollaboration2020flux} and compares the peak fluxes of three flares possessing an X-ray counterpart. The flare under investigation here is almost an order of magnitude fainter and a factor of $\sim2-3$ shorter than previously analysed very bright X-ray flares \citep{Dodds-Eden2009, Ponti2017}.
Thanks to the frequent observations of \sgras's X-ray emission, more than a hundred X-ray flares of \sgras\ have been detected so far by \chandra\ and \xmm\  \citep{Neilsen2013, Ponti2015, 2016A&A...589A.116M, Li2017, Bouffard2019}). Figure \ref{fig:fluence} highlights the fluence and duration of the X-ray flare detected here and compared to previously detected flares. 

The July 18 flare shows only moderate emission in the X-ray band. Indeed, it is almost an order of magnitude fainter and a factor of $\sim2-3$ shorter than the very bright X-ray flares for which the IR to X-ray spectral energy distribution has been investigated in detail in previous works \citep{Dodds-Eden2009, Ponti2017}. The relative X-ray faintness is unexpected, considered that the flare is one of the brightest flares in the IR band. 

\section{Analysis of the mean spectrum\label{sec:mean spectrum}}
\subsection{IR spectrum}
\begin{figure}
    \centering
    \includegraphics[width=0.45\textwidth]{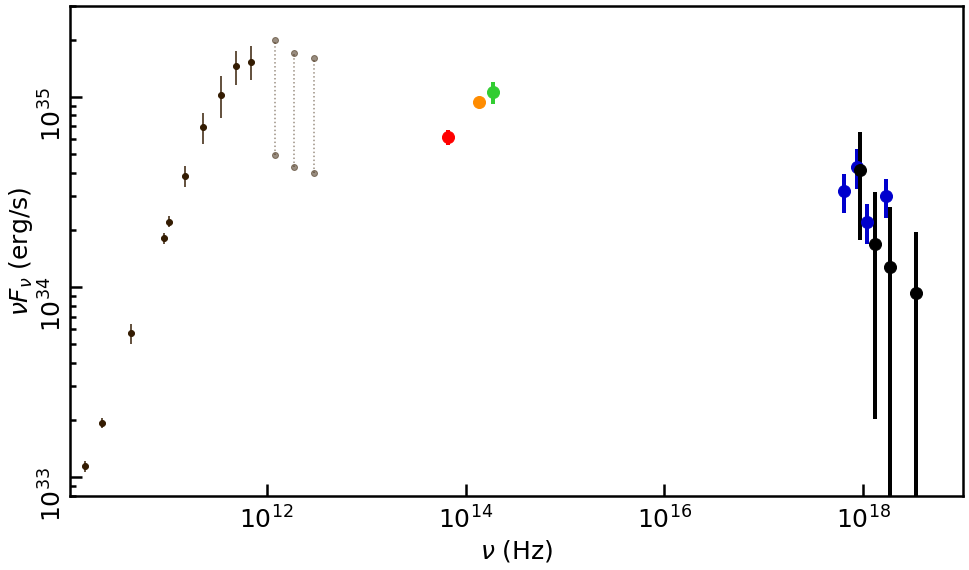}
    \caption{The mean SED plotted together with the best fit power law slope. The sub-mm SED is plotted for orientation; the radio and sub-mm data are from \cite{Falcke1998, Bower2015, Brinkerink2015, Liu2016, Bower2019}. The far infrared data are from \cite{Stone2016} and \cite{VonFellenberg2018}.}
    \label{fig:SED}
\end{figure}

In order to obtain the mean spectrum, we binned all six exposures with significant IR flux to find the average flux density in the $M$-, $K$- and $H$-bands. These flux densities were converted to luminosities and are shown in \autoref{fig:SED}. 

\subsection{\textbf{\textit{Chandra}}} 

\begin{figure} 
    \centering
    \includegraphics[width=0.45\textwidth]{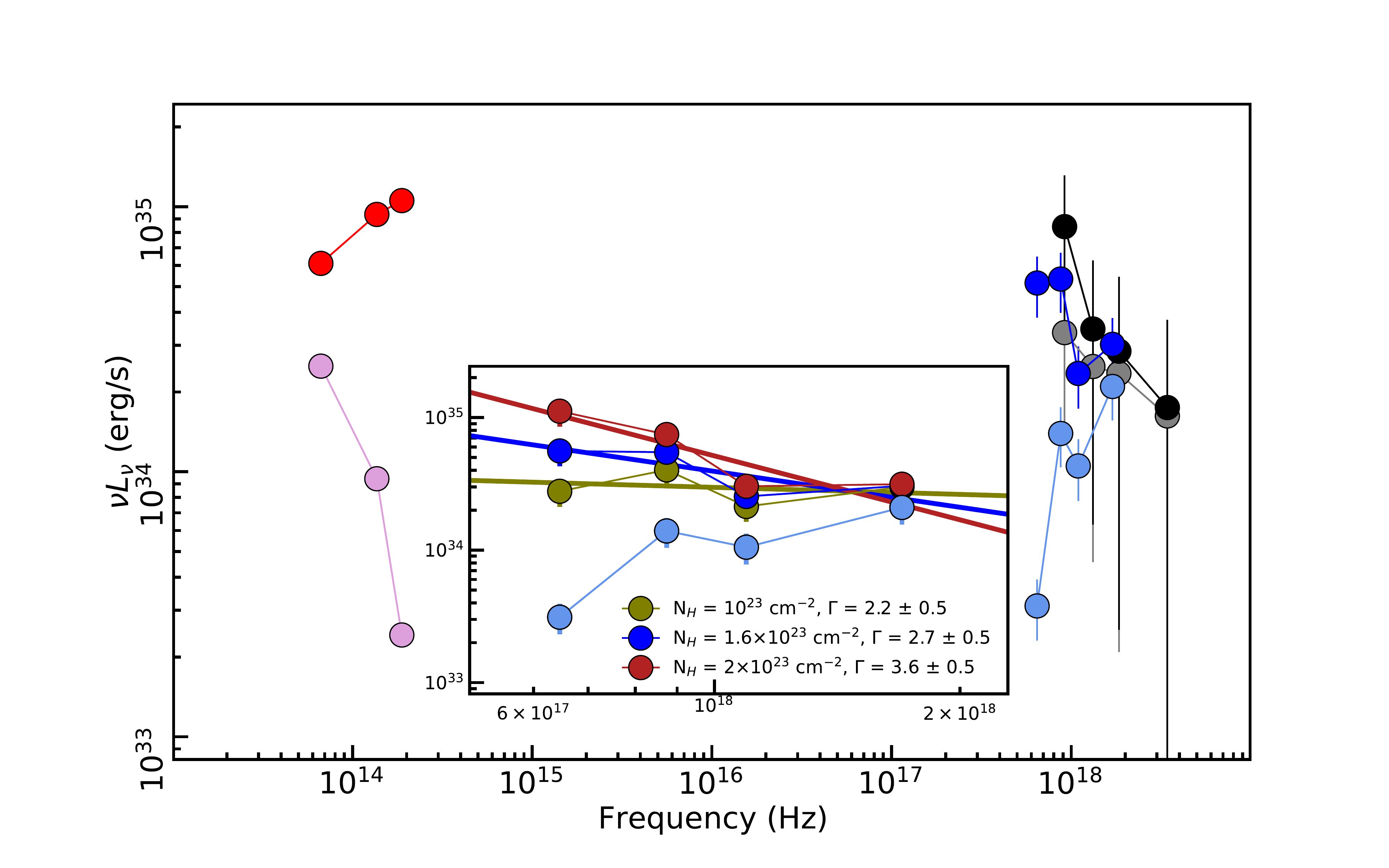}
    \caption{{\it Main panel:} Comparison between observed and corrected spectra. The cyan, grey, and pink points show the spectra as observed by \chandra, \nustar, and in the IR band, respectively. The blue, black, and red points show the same data corrected for absorption and the effects of dust scattering. The correction amounts to more than one order of magnitude in K and H as well as in the soft X-ray band. {\it Inset: } As in the main panel, the cyan points show the spectrum as observed by \chandra. The olive, blue, and dark red points show the \chandra\ spectrum after correction assuming $N_{\rm{H}}=10^{23}$, $1.6\times10^{23}$, and $2\times10^{23} \mathrm{cm^{-2}}$, respectively.}
    \label{DeAb}
\end{figure}

Dust extinction and absorption due to neutral material along the line of sight are a major source of systematic uncertainty for all observations of the Galactic Center. A fit of the original \chandra\ spectrum with an absorbed power law, corrected for the distortions introduced by dust scattering, provides a best fit photon index $\Gamma=2.7\pm0.5$ (C-stat = 238.1 for 545 dof). 
The best-fit 2--10~keV observed flux is $F_{\rm{Abs}~2-10}=2.3\times10^{-12}$ erg cm$^{-2}$ s$^{-1}$. Once de-absorbed and corrected for the effects of dust scattering, this corresponds to $F_{\rm{Deabs}~2-10}=6.9\times10^{-12}$ erg cm$^{-2}$ s$^{-1}$. 
In order to fit the temporal evolution of the spectrum together with the NIR data, we rebinned the observed spectrum to have 4 bins in energy each containing 18 photons. For the time-resolved spectra we binned our spectra in 2, 2, and 1 bins containing 16, 14, and 12 photons for T2, T3, and T4, respectively. 
Starting from the best-fit model of the original data, we computed the ratio between the absorbed and scattered model and the de-absorbed and dust-scattering-corrected model. We then applied this model ratio to the rebinned spectrum to derive the corrected spectrum of \sgras's flare. 

The effects of absorption and dust scattering are very significant in the soft X-ray band. Indeed, a comparison between the observed and de-absorbed spectra shown in Fig. \ref{DeAb} shows a ratio in excess of one order of magnitude below ${\sim}3~\mathrm{keV}$. The soft X-ray flux and X-ray photon index are strongly correlated dependent on the assumed column density of absorbing material (see of \autoref{DeAb}). By assuming column densities of $N_H=10^{23}$, $1.6\times10^{23}$ and $2\times10^{23}$~cm$^{-2}$ (all values which are consistent with the spectrum of this moderate X-ray flare), the best fit photon index is $\Gamma=2.2\pm0.5$, $2.7\pm0.5$, and $3.6\pm0.5$, respectively. These values are consistent with the allowed range of values reported in works compiling several X-ray flares (e.g. \citealt{Porquet2008, Nowak2012}). To allow a better comparison with previous multi-wavelength flares of \sgras, we assume $N_H=1.6\times10^{23}$~cm$^{-2}$ \citep{Ponti2017}. We discuss the implications of this choice in Appendix \ref{appendix:nh_density}.

\subsection{\textbf{\textit{NuSTAR}}}

As a consequence of the larger point-spread function of the \nustar\ mirrors, a larger fraction of diffuse emission contaminates the \nustar\ spectra of \sgras compared to \chandra. 
Indeed, \sgras's photons amount to about 30~\% of the total flux in the 3--20~keV band. In order to reduce the uncertainties associated with background subtraction, we fitted the background spectrum simultaneously with the source plus background, adopting the same background model components in both cases.

We parametrised the \nustar\ background spectrum in the 3--50~keV band with a collisionally-ionised diffuse plasma component ({\sc apec} in {\sc xspec}) plus a power law, all absorbed by neutral material. 
This model provides a good description of the background spectrum (see Tab. \ref{fitX}). 
We simultaneously fitted the source plus background spectrum by adding an absorbed power-law component to this model to fit the emission from \sgras. 
The best fit photon index of \sgras's emission is $\Gamma=2.6\pm1.0$ with an absorbed 3--20 keV flux of $F_{\rm{Abs}~3-20}=3.1\times10^{-12}$ erg cm$^{-2}$ s$^{-1}$ ($F_{\rm{Deabs}~3-20}=4.5\times10^{-12}$ erg cm$^{-2}$ s$^{-1}$). 
\begin{table}
\begin{center}
\begin{tabular}{ l c c c c }
\hline
\hline
\multicolumn{4}{c}{\bf X-ray spectral analysis} \\
& \chandra & \nustar & \chandra+ \\ 
&  & & \nustar \\ 
\hline
\hline
{\bf \sgras} \\
$\Gamma$   & $2.7\pm0.5$      & $2.6\pm1.0$       & $2.7\pm0.5$ \\
$N_{pl}$   & $87^{+90}_{-45}$ & $50^{+300}_{-40}$ & $67^{+90}_{-40}$ \\
{\bf Background} \\
$kT_{a}$     &                & $1.8\pm0.2$       & $1.8\pm0.2$ \\
$N_{H}$      &                & $2.4\pm0.4$       & $2.6\pm0.4$ \\
$\Gamma$     &                & $1.7\pm0.1$       & $1.7\pm0.1$ \\
$N_{pl}$     &                & $12\pm4$          & $13\pm4$    \\
\hline
C-S/dof      & 238.1/547      & 1046.6/1717       & 1284.6/2264\\
\hline
\hline
\end{tabular} 
\caption{Parameters of the best fit to the \chandra, \nustar, and combined 
source and background spectra. 
$N_H$: column density of neutral material ($10^{22}$ atoms cm$^{-2}$);
$\Gamma$: photon index of power law component; 
$N_{pl}$ normalisation ($10^{-4}$ photons keV$^{-1}$ cm$^{-2}$ s$^{-1}$ at 1 keV) of the power law component; 
$kT_{a}$ plasma temperature (keV) of the {\sc apec} component; 
$N_{a}$ normalisation ($10^{-2}$) of the {\sc apec} component; 
C-S: value of Cash statistic.}
\label{fitX}
\end{center}
\end{table} 

\subsection{Combined fit of \textit{Chandra+NuSTAR} spectra}

Finally, we simultaneously fitted the background subtracted \chandra\ as well as the source+background and background \nustar\ spectra. This provides a good fit to the data, with a best-fit $\Gamma=2.7\pm0.5$ (see Tab. \ref{fitX}). 

To perform multi-wavelength fits with models not yet implemented in {\sc xspec} (e.g.\ synchrotron cooling break and high energy cutoff SSC models), we corrected the binned \chandra\ and the binned\footnote{The \nustar\ spectrum has been rebinned in order to have 21 photons per bin in the 3--40~keV energy band.} background-subtracted \nustar\ spectrum for the effects of absorption and dust scattering and then fit the corrected spectrum with a least squares fit. This step might introduce biases in the corrected spectrum. However, we verified that such distortions are negligible compared to the statistical uncertainties of the X-ray spectra.

\section{Temporal evolution of the SED\label{sec:temporal spectrum}}
\begin{figure}
    \centering
    \includegraphics[width=0.45\textwidth]{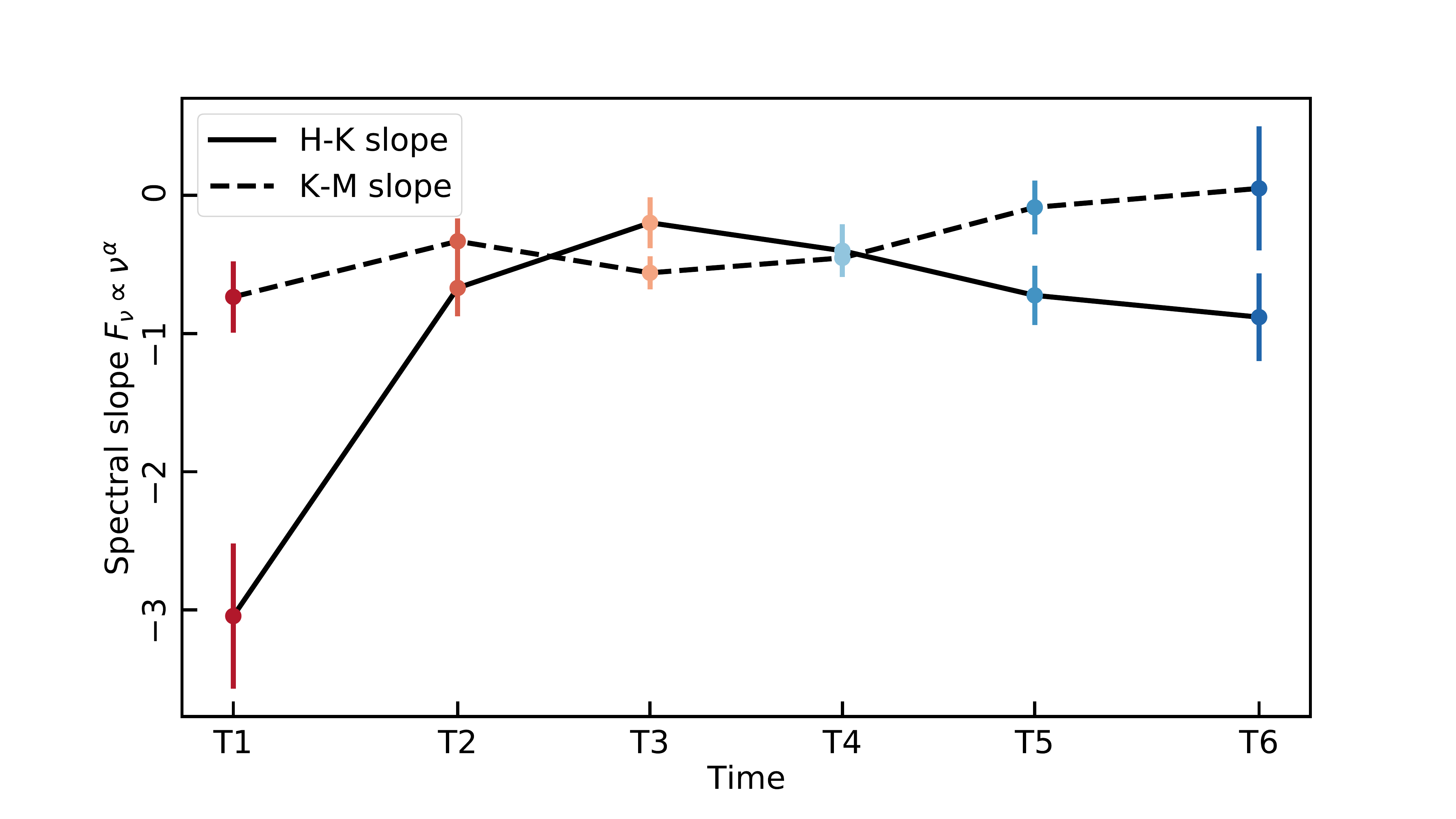}
    \caption{The IR spectral slopes $\alpha$ for the six times T1 to T6. The color encodes the time, dark red to dark blue. The black solid line shows the $H-K$ slope; the black dashed line shows the $K-M$ slope. }
    \label{fig:nir spectrum}
\end{figure}

We can determine a spectral index for each of the six exposures with significant IR flux,. Here we report the spectral slope of the flux density $F_\nu \propto \nu^{\alpha}$. The spectral slope of the luminosity is $\nu F_{\nu} \propto \nu^{\beta}$, where $\beta=\alpha + 1$. In order to compare the spectrum of the $M$-band to the $K$- band and the $K$-band to the $H$-band, we analytically computed the spectral slope:
\begin{equation}
    \alpha_{\rm Band1 - Band2} = \log_{(F_{\rm Band1}/ F_{\rm Band2})}(\nu_{\rm Band1}/\nu_{\rm Band2}).
\end{equation}
and propagated the uncertainty of the observed flux densities (\autoref{fig:nir spectrum}).

\begin{figure*}
    \centering
    \includegraphics[width=0.95\textwidth]{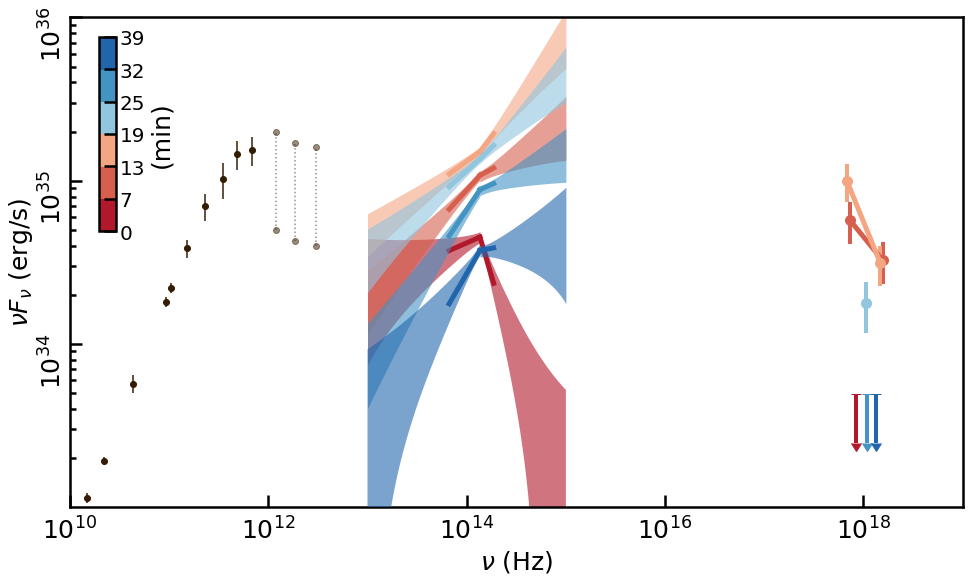}
    \caption{The temporal evolution of the SED. The color encodes the time: dark red to dark blue as indicated in the color bar. For two time steps, the X-ray spectrum can be split up into two points (T2 and T3). For T4, only one X-ray flux measurement is possible. For T1, T5, and T6 upper limits are plotted. The measurements in the NIR are indicated by thick lines, with the uncertainties indicated and extrapolated by the shaded area. The sub-mm data shown are the same as in Figure \ref{fig:SED}.}
    \label{fig:temporal SED}
\end{figure*}

During the onset of the flare, \sgras{} was faint in the $H$-band, while there is already substantial flux measured in the $M$- and $K$-bands. This resulted in a very red $H-K$ slope ${\sim} -3$, while the $K-M$ slope was ${\sim} -0.7$. After the first data point, the $H-K$ slope jumped to ${\sim} -1$. For the next two data points, the spectral slope increased from $\alpha_{H-K}{\sim} - 1$ to $\alpha_{H-K} {\sim} 0$ at the peak of the flare. After the peak $\alpha_{H-K}$ decreased, with $\alpha_{H-K}{\sim} -1$ at the end of the flare. This indicates a correlation between the $H-K$ spectral slope and the flux density. Conversely, there was no strict correlation of the spectral slope with flux density for $\alpha_{H-K}$. The $K-M$ slope varied in the range $\alpha_{K-M} = [-0.8, 0.0]$ and increased towards the end of the flare. However, this might be indicative of a correlated error due to a telescope slew of the \spitzer{} spacecraft. The temporal evolution of the flare SED is shown in \autoref{fig:temporal SED}.

\section{One Zone SED model}
To model the IR to X-ray SED of \sgras, we developed a dedicated python package (Dallilar et al. in prep.). 
The code implements robust calculation of synchrotron emission or inverse Compton scattering from a given underlying electron distribution in a single zone. We also provide a convenient SED fitting interface built on top of the general purpose python fitting package LMFIT\footnote{\url{https://github.com/lmfit/lmfit-py/}}.
For testing and convenience, the code includes theoretical solutions to synchrotron emission and absorption coefficients of a thermal, power law, or kappa distribution based on the formalism presented by \cite{Pandya2016}. Furthermore, we implemented a fast numerical calculation of the emission and absorption coefficients for a given arbitrary electron distribution. With this feature, we are able to explore more complex electron distributions. This is especially important in the context of including ``cooling break" types of models \citep{Dodds-Eden2009, Ponti2017} and more realistic cutoff shapes of the electron distribution. Our approach is an improvement compared to similar attempts in the aforementioned works in terms of self-consistent determination of electron distribution parameters from SED fitting. The code's inverse Compton scattering formalism follows the concepts presented by \cite{Dodds-Eden2009}. As with synchrotron emission, we can take advantage of arbitrary electron distributions as the scattering medium. Seed photons can be either an external (arbitrary) photon field or synchrotron emission from an underlying electron population, namely, synchrotron self-Compton emission. The details of the code will discussed by Dallilar et al., (in prep.)

The philosophy of the code is to provide emission scenarios that are as simple as possible. This is achieved by modeling the flares in a scenario in which the emission is dominated by a single localized region in the accretion flow and by a single population of electrons, reducing the number of free parameters. For instance, if the emission is modeled using a power-law distribution of electrons, the number of free parameters is six. Keeping the number of free parameters small is necessary because our limited spectral coverage does not warrant a more complex fit (i.e the number of model parameters should be smaller than the number of observables). Therefore, the luminosity is computed for a homogeneous and spherical geometry of electrons. Ultimately, we can fit the model SED to the data, either through $\chi^2$ minimisation or through MCMC modeling.

\section{Reproducing the mean SED of the flare}

\subsection{Synchrotron with a cooling break}
\label{sec:plc model}
\begin{table*}
\begin{center}
\begin{tabular}{ l c c c c c c c c c }
\hline
\hline
 & \multicolumn{3}{c}{\bf Mean SED} & \multicolumn{6}{c}{\bf Time Resolved} \\
 & PLCool & \plcg$_{\rm{sharp}}$& \plcg\ & T1 & T2 & T3 & T4 & T5 & T6 \\
\hline
$\log(n_{e}\times~1\mathrm{cm^{-3})}$& 6.7$\pm$0.2& 6.3$\pm$0.2 & $5.52\pm0.01$& $5.5\pm 0.2$&$5.6\pm0.1$&$5.8\pm0.1$ &$5.8\pm0.1$   &$5.7\pm 0.1$&$5.3\pm0.1$ \\
$R~[\mathrm{R_S}]$               & 1\dag      & 1\dag       & 1\dag     &1\dag       &1\dag      &1\dag        &1\dag          &1\dag      &1\dag   \\
$B~[\mathrm{G}]$               & 38$\pm$6 & 30\dag    & 30\dag  &30\dag      &30\dag     &30\dag       &30\dag         &30\dag     &30\dag  \\
$p$               & 2.4$\pm$0.1& 2.0$\pm$0.1 & 2\dag     &2\dag       &2\dag      &2\dag        &2\dag          &2\dag      &2\dag   \\
$\gamma_{max}$    & $>10^{3}$  & 68$\pm$13    & $48\pm4$  &$1.5\pm1.4$ &$52\pm0.7$ &$43\pm5$     &$29\pm4$       &$5$\dag    &$5$\dag \\
$\chi_{red}^2$ / DOF& 5.0 / 3       & 2.2 / 4        & 1.1 / 2      & 4.9 / 2     & 0.6 / 2     & 0.8 / 2     & 0.7 / 2       & 5.7 / 1    & 2.0 / 1 \\
\hline
\hline
\end{tabular} 
\caption{Best fit parameters of the fit of the SED with the PLCoolgamma model. 
$n_{e}$: electron density within the source; 
$p$: power law index of the electron distribution; 
$R$: projected radius, in Schwarzschild radii, of the emitting source; 
$B$: magnetic field intensity (G); 
$\gamma_{max}$: maximum Lorentz factor of the accelerated electrons in units of $10^3$; 
$\chi^2_{red}$; DOF: reduced $\chi^2$ of the best fit, number of free parameters 
\dag: value fixed. The uncertainties reported correspond to the $1\sigma$ confidence limits determined through MCMC sampling.
}
\label{tab:PLCoolgamma}
\end{center}
\end{table*} 

\begin{figure*}
\centering
\includegraphics[width=0.95\textwidth,angle=-0]{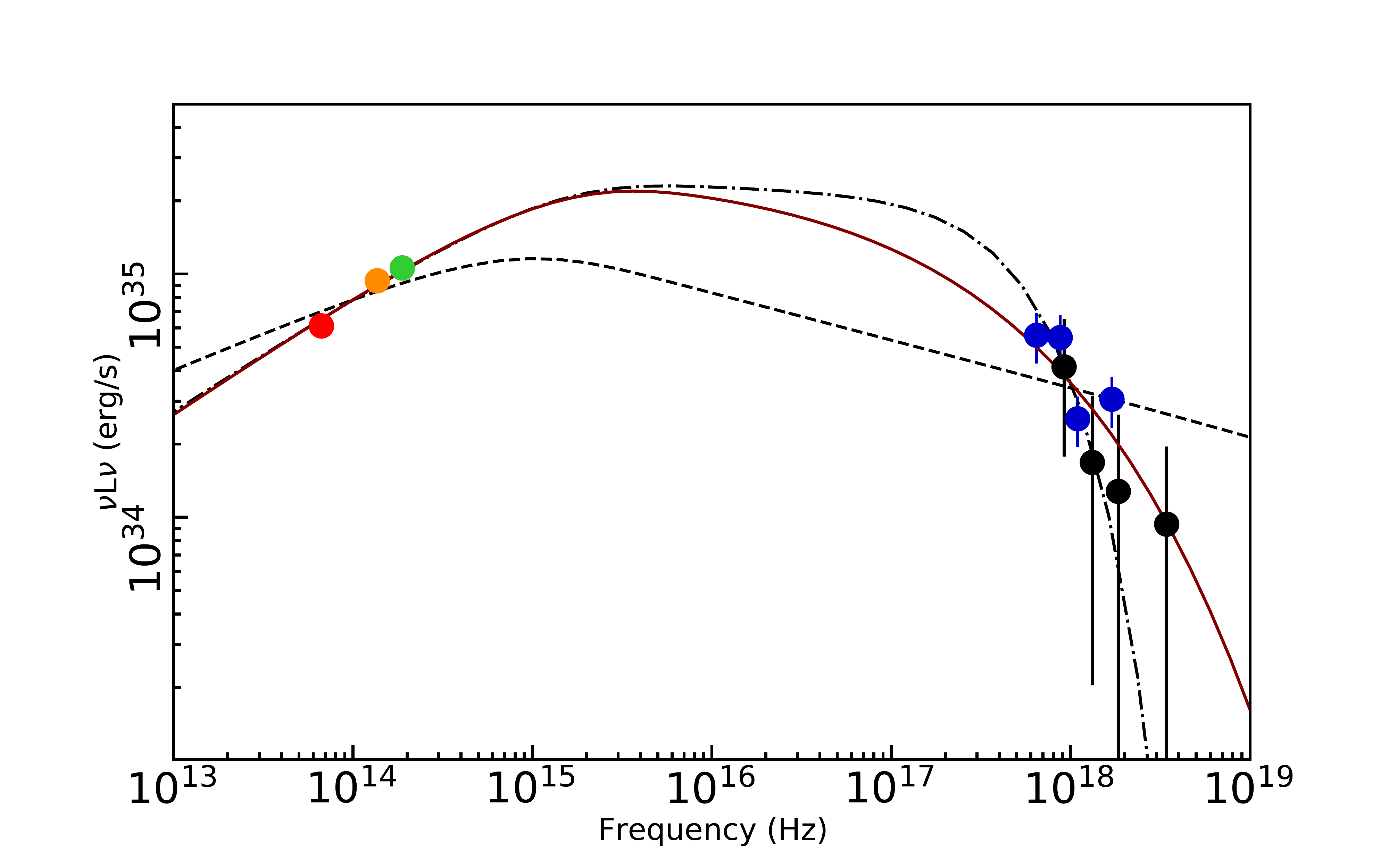}
\caption{Mean SED of Sgr~A* during the flare as in \autoref{fig:SED}, including the best fit synchrotron models. The black dashed line shows the best-fit PLCool model (synchrotron with cooling break model with no high energy cutoff). 
This model is ruled out because it cannot fit the difference in X-ray vs.\ IR spectral slopes due to the X-ray vs.\ IR flux ratio. 
The dashed-dotted black lines shows the best fit \plcg$_{\rm{sharp}}$ model (synchrotron with cooling break plus a sharp $\gamma_{max}$ cutoff). The line cuts off too sharply in the X-ray and fails to reproduce the high energy \nustar{} data. 
The dark red line shows the best fit \plcg{} model (synchrotron with cooling break plus an exponential high energy cutoff). For this model, we have also computed the synchrotron self-Compton component, which peaks at $\nu\sim 10^{23}~\mathrm{Hz}$ (not shown here, see \autoref{fig:MeanSEDPLSSC}).}
\label{MeanSEDPL}
\end{figure*}
We began by fitting the mean spectrum of the flare with a simple synchrotron model with a cooling break (see Fig. \ref{MeanSEDPL}). We call this model the PLCool model.
Although the difference in photon indices between the IR ($\Gamma_{\rm{IR}}=1.5\pm0.2$) and X-ray ($\Gamma_{\rm{X}}=2.7\pm0.5$) bands is consistent with the expectations of the synchrotron model with cooling break ($\Delta\Gamma=0.5$), it is not possible to fit the mean SED of the flare with this model. Indeed, the high luminosity in the IR band combined with the rather flat IR spectrum would imply a very high luminosity in the X-ray band. As a consequence of this tension, the PLCool model settles to a less blue IR slope than observed, failing to satisfactorily fit the data (Figure \ref{MeanSEDPL} and Table \ref{tab:PLCoolgamma}). 

\subsection{Synchrotron with a cooling break and sharp high energy cutoff}
\label{sec:plcg model}
The acceleration mechanism generating the flare may not be powerful enough to accelerate particles to $\gamma_{max}\gg10^5$ at all times \citep{Ponti2017}. If this is true, we expect to observe a high-energy cutoff between the IR and X-ray bands. 
Hence, we fit the mean SED with a synchrotron model with cooling break and a high-energy cutoff in the electron distribution. We call this model the \plcg$_{\rm{sharp}}$.
In particular, we assumed that the high energy cutoff is a step function with no electrons having $\gamma>\gamma_{max}$. 
We assumed that the electrons are accelerated from the thermal pool which is producing the sub-mm emission, and therefore we fixed $\gamma_{min}=50$.  
We assume a source with $1 ~\mathrm{R_s}$ radius, a magnetic field strength of $B=30~\mathrm{G}$ and a cooling time of 2 minutes (Tab. \ref{tab:PLCoolgamma}). 
A fixed cooling time scale of 2 minutes was motivated by the light travel time for a source with radius $1 ~\mathrm{R_s}$: the cooling-break model assumes an equilibrium of particle acceleration and particle losses due to particle escape, and thus particles at low energy escape the flare region before they cool. In consequence, the position of the cooling break in the spectrum corresponds to the electron energy at which the escape time is equal to the cooling time \citep{Kardashev1962, Yuan2003}. Following \cite{Dodds-Eden2009}, we assume that the escape from the system can be approximated by the dynamical time scale. This assumption, together with our assumption of a magnetic field strength of $B=30~\mathrm{G}$, fixes the cooling break:
\begin{equation}
\begin{aligned}
       \nu_B &= 64\cdot(B/30\mathrm{[G]})^{-3}\times10^{14}/t_{cool}^2 \\
             &= 1.6\times 10^{15}~\mathrm{Hz}.
\end{aligned}
\end{equation}

This model provides a decent description of the data with acceptable physical parameters, as shown in Fig. \ref{MeanSEDPL}. The best-fit $\log(n_e)=6.3\pm0.2$ and slope of the electron distribution, $p=2.0\pm0.1$ are in line with the density expected in \sgras's hot accretion flow and the electron distribution undergoing synchrotron cooling $p\geq2$ \citep{Kardashev1962, Ghisellini2013}. 
On the other hand, the model predicts a significantly softer X-ray emission than observed. Large residuals are observed at high energy, where the model decays quickly with frequency, while the data indicate a clear excess of emission associated with the flare all the way from ${\sim} 2$ to ${\sim}8$ keV. 
Therefore, this model is also unsatisfactory. 

\subsection{Synchrotron with a cooling break and exponential high energy cutoff}

A more realistic model is an exponential decay of the electron distribution above a certain cutoff energy. This induces a shallower spectrum at high energy. A synchrotron model with a cooling break and exponential high-energy cutoff can fit the data in an acceptable way. We call this model the \plcg.
The slope of the electron distribution is $p=2.0\pm0.2$, consistent with the cooling break scenario \citep{Kardashev1962,Ghisellini2013}. 
The density $n_e=10^{5.5\pm0.1}~\mathrm{cm^{-3}}$ of accelerated electrons suggests that only a fraction of the electrons in the hot accretion flow are involved in the acceleration process. 
Finally, the best-fit $\gamma_{max}=(4.8\pm4.0)\times10^4$ induces a cutoff in the X-ray band explaining the observed X-ray faintness. 

\section{Time-resolved evolution of \sgras's SED during the flare}
\subsection{Synchrotron with a cooling break and high energy cutoff}
\begin{figure}
    \centering
    \includegraphics[width=0.45\textwidth]{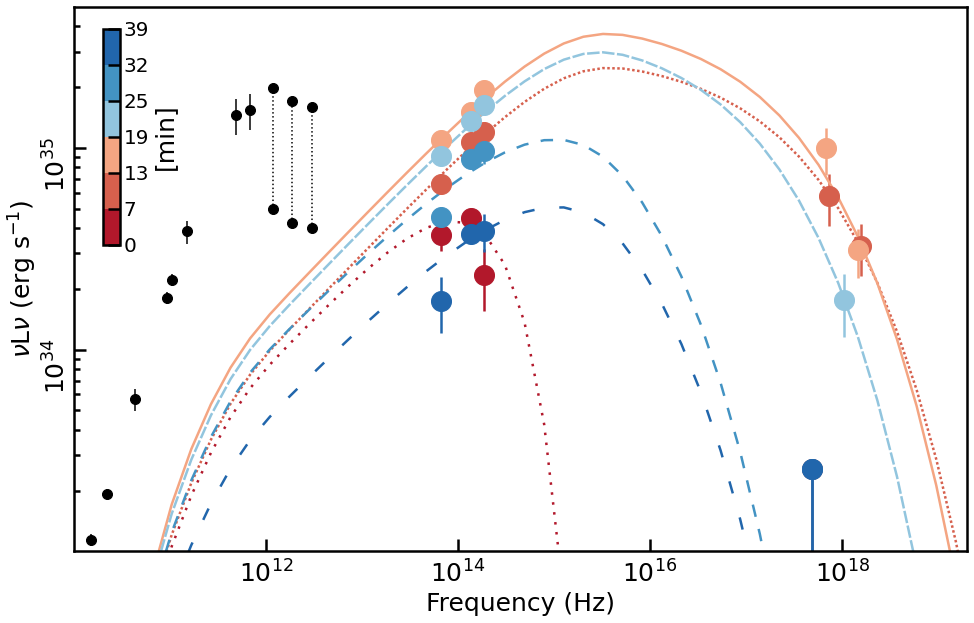}
    \caption{The data points show the \spitzer\ + \gravity, and \chandra\ photometry during T1 to T6, respectively (dark red to dark blue lines). The data are corrected for the effects of absorption and dust scattering. The lines show the best-fit synchrotron with cooling break and high energy cutoff models. During the early phases of the flare, the high energy cutoff appears to be at low energy. During the peak of the flare, the cutoff moves to the X-ray band and then drops again to low energies towards the end of the flare. The sub-mm data shown are the same as in Figure \ref{fig:SED}, and the color-bar indicates the time/color progression.}
    \label{fig:TRSED}
\end{figure}

\autoref{fig:TRSED} shows \sgras's SED temporal evolution during the flare fitted with the \plcg\ model. Table \ref{tab:PLCoolgamma} reports the maximum likelihood fit parameters and their uncertainties from the $1 \sigma$ posterior contours of the MCMC sampling. 

For T1, T5, and T6, no X-ray flux was detected. For these three time steps, therefore, the spectrum is composed of only three data points. For T2 and T3, significant X-ray flux was observed, which allows us to determine the flux of \sgras\ in two energy bins. For T4, we binned the high-energy band to one data point. Because of the limited number of free parameters in this time-resolved analysis and in the interest of reducing the number of free parameters in our model, we fixed the magnetic field strength $B$ and the source radius $R$ to $B=30~\mathrm{G}$ and $R=1~\mathrm{R_S}$. 
However, we left the particle density $n_e$ free. The particle density primarily drives the normalisation of the spectrum. 
The magnetic field strength, the radius, and the particle density are degenerate in the model. Therefore an error in our assumed values of the magnetic field strength and source radius would be compensated by the electron density. 

We did not attempt to model the evolution of the electron distribution self-consistently. This would require assuming an emission zone expansion, an electron injection, and an electron cooling scenario. While informative, such scenarios have been explored in one-zone models of flares before (e.g. \cite{Dodds-Eden2010, Dibi2014}) and we assume that the conclusions found in these studies are applicable. The analysis of the mean SED of this flare requires a maximum acceleration $\gamma_{max}\sim 10^4$, and we focused our modeling on the evolution of this parameter.

The minimum acceleration of the electrons is based on the sub-mm emission and fixed at $\gamma_{min}=50$. Motivated by the fit to the mean SED, we fixed the slope of the electron distribution to $p=2$. Therefore, the free parameters in the model are $n_e$ and $\gamma_{max}$. Fixing the electron distribution slope precludes the possibility to explore the changes of spectral slope shown in \autoref{fig:nir spectrum}. These choices and assuming that the cooling time scale is set by the escape time of particles escaping the emission region fixes the cooling break at $\nu =1.6\times 10^{15}~\mathrm{Hz}$.

At the start of the flare during T1 (\autoref{fig:TRSED}), relatively bright emission was observed in the $M$- and $K$-bands, while fainter emission was observed in the $H$-band and no excess emission was detected in the X-ray band. 
If the IR emission is produced by non-thermal synchrotron emission with a positive IR spectral slope (in $\nu F_\nu$), then the lack of X-ray emission implies that the distribution of relativistic electrons must have a cutoff at high energy. The flare was bright in the $M$- and $K$-bands during T1, but it was barely detected in the $H$-band which can be understood in the framework of the \plcg{} model. If the maximum acceleration of the electrons ($\gamma_{max}$) happens to be located within the $K$- or $H$-band, then the flux drops in the $H$-band and no X-ray emission is expected, in line with the observational results. However, this does not explain the drop in flux between the $K$- and $H$-bands. The \plcg{} model only marginally matches the data, with the $H$-band flux being too faint compared to the $K$- and $M$-bands. This might be a consequence of an underestimated uncertainty for the marginal $H$-band detection. 

In T2, the IR flux increases and the slope was consistent with a power law from the $M$- to the $H$-band, and significant X-ray flux was detected. The data are well-fit by the \plcg{} model, with the maximum acceleration at frequencies slightly higher than the X-ray band. For T2 (shown by the red SED in \autoref{fig:TRSED}), the fitted acceleration reaches its maximal value $\gamma_{max}=(52\pm1)\times10^3$.
During the following interval (T3, shown by the light red SED in \autoref{fig:TRSED}), both the IR and X-ray emission are at their peaks. However, although little variation in the spectral slope was observed in the IR band, the simultaneous X-ray spectrum appears softer. Our model ascribes this to the maximum acceleration of the electrons having decreased to $\gamma_{max}=(43\pm5)\times10^3$. 
Subsequently, in T4, the flux starts to drop in both bands (shown by the light blue SED in \autoref{fig:TRSED}). However, although the drop in the IR band is of the order ${\sim}20$~\% (\autoref{fig:TRSED} and \autoref{tab:PLCoolgamma}), again with little variation in the spectral slope, the flux in the X-ray band dropped by more than a factor of 3. Within the framework of the \plcg\ model, this can be ascribed to the acceleration mechanism continuing to lose the capability to accelerate electrons to the highest energies, therefore moving $\gamma_{max}$ to $(29\pm4)\times10^3$. This puts the high energy cutoff in the electron distribution between the IR and X-ray bands, and the X-ray emission at this time would be produced mainly by electrons above the cutoff. 

No X-ray emission was observed during the subsequent intervals T5 and T6 (shown by the blue and dark blue in \autoref{fig:TRSED}). As in T1, the IR was still bright (${\sim}5-10$~mJy) and flat.
The \plcg\ model reproduces this by placing the high energy cutoff somewhere between the IR and X-ray band. We thus obtain an upper limit on $\gamma_{max} < 5000$.

During these last two intervals, the $M$-band flux dropped faster than the $K$- and $H$-band fluxes. This resulted in a blue $K-M$ slope, which would imply a decrease of $p$ to $p\sim1.4$, while the $H-K$ slope was consistent with $p\sim2$. If taken at face value, the observed $M$-band flux was inconsistent with a fixed slope of $p=2$ and is responsible for the worse $\chi^2$ for T5 and T6. However, this may be attributable to a correlated error in the relative flux measurement, due to a telescope slew (Section \ref{sec:temporal spectrum})

\subsection{Temporal evolution of the electron distribution}
\begin{figure}
    \centering
    \includegraphics[width=0.45\textwidth]{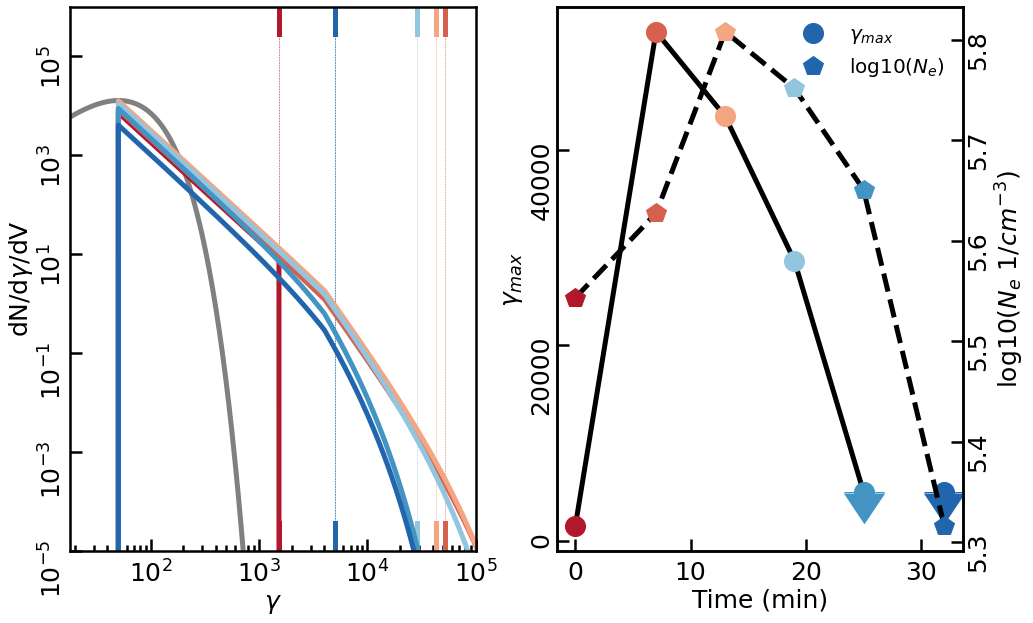}
    \caption{Left: The evolution of the electron distribution during the flare.  The different temporal steps are plotted dark red (T1), progressing to lighter reds (T3), to light blue (T4) to dark blue (T6). The dotted lines indicate the location of $\gamma_{max}$. The grey line shows a thermal distribution of electrons, peaking at $\gamma\sim 50$, which set the minimum acceleration of the electrons for the flare. Right: evolution of the distribution parameters $\gamma_{max}$ (shown by the solid line) and $n_e$ (shown by the dashed line).}
    \label{fig:electron_dist_1col}
\end{figure}

\autoref{fig:electron_dist_1col} reports the energy distribution of the accelerated electrons for each of the time bins during the flare. It also we shows the energy distribution of the electrons responsible for the sub-mm emission. In order to match the sub-mm SED of \sgras{}, we computed the spectrum assuming values within the range of parameters reported by \cite{Bower2019}. For the sub-mm emission, we assumed an ambient magnetic field strength $B=30~\mathrm{G}$, as for the flare, and a size of $4~R_S$, consistent with the observed sub-mm size \citep{Issaoun2019}. We chose an ambient particle density $\log(n_e) = 1.7\times 10^5$ such that the distribution peaks at $\gamma_{min}=50$.
The right panel of \autoref{fig:electron_dist_1col} shows that within 380~s, $\gamma_{max}$ reaches its maximum value of $\gamma_{max} \sim 5\times10^4$, indicating that the most energetic electrons are accelerated during T2. In the following intervals, the maximum $\Gamma$ steadily decreases, and we can only constrain it to values below $4\times 10^3$ once the X-ray flux has dropped below the detection limit. The electron density, plotted in the right panel of \autoref{fig:electron_dist_1col}, reaches its maximum when the flux is the highest (T3), after which it steadily decreases.

\begin{figure*}
    \centering
    \includegraphics[width=0.95\textwidth,angle=-0]{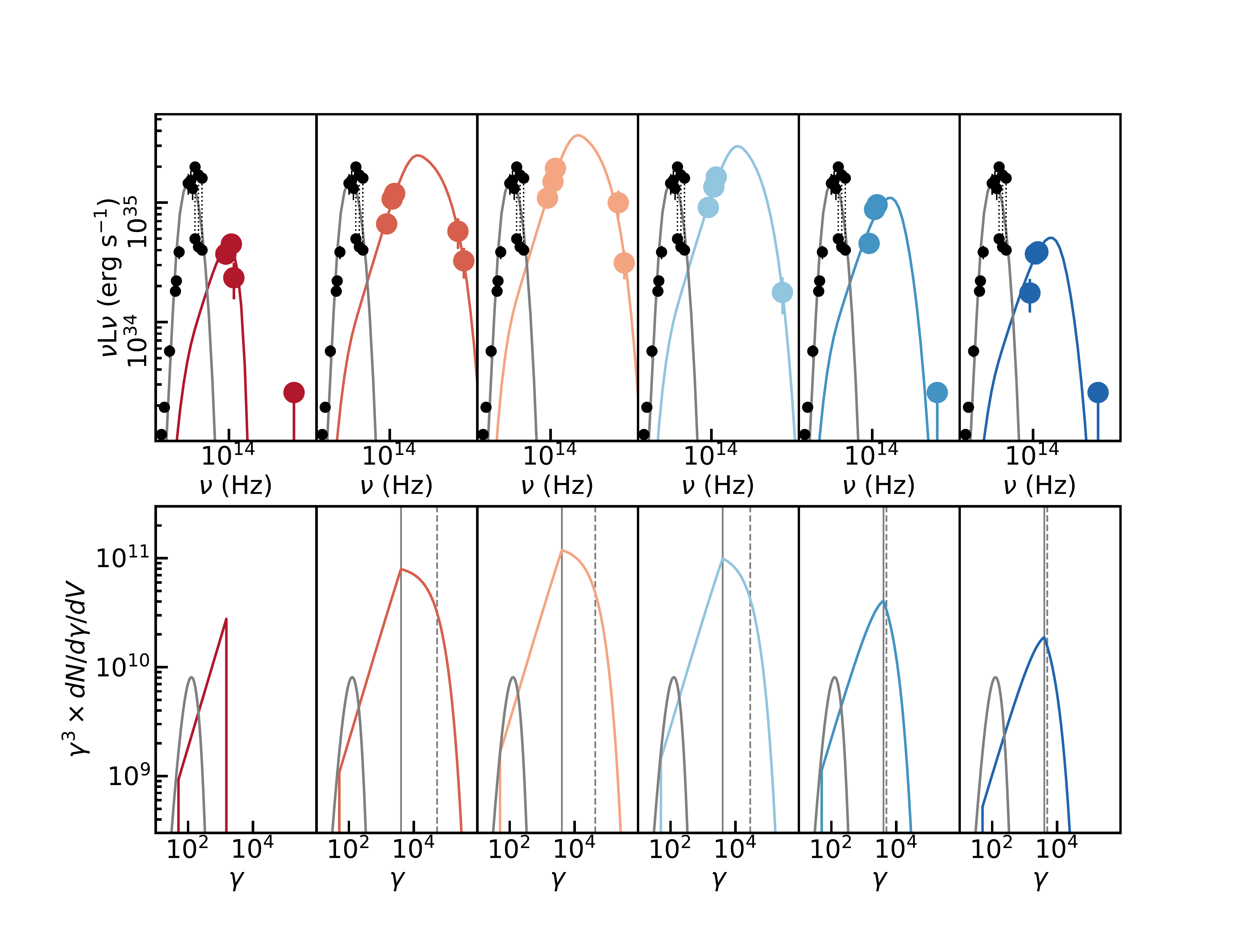}
    \caption{The temporal evolution of the flare SED and the temporal evolution of the electron energy distribution. Panels left to right show the temporal evolution from T1 to T6. 
    Top row: the observed SED of the flare (colored points) and the best-fit \plcg{} model (colored lines). The black points indicate the sub-mm SED of Sgr~A*, with the same data as in Figure \ref{fig:SED}. The thin grey line shows a thermal synchrotron spectrum matching the sub-mm data. 
    Bottom row: the electron energy distribution of the respective synchrotron spectra in the top row. Colored lines show the best-fit \plcg{} models, the thin grey line shows the electron energy distribution of the thermal spectrum. The positions of the cooling break and $\gamma_{max}$ are indicated with solid and dashed grey lines, respectively. To highlight the location of the breaks in the distributions, the cooling break, and the maximum acceleration we have multiplied the electron distribution by a factor $\gamma^3$.}
    \label{fig:TRSED_all}
\end{figure*}

\autoref{fig:TRSED_all} shows the evolution of the time-resolved SED fitted with the \plcg model along with the respective electron distributions as inferred from the best fit. 

\subsection{An alternative model: synchrotron self-Compton scattering of sub-mm photons}

An alternative scenario to explain the temporal evolution of Sgr~A* variability has been proposed by Witzel et al. (2021). Using a comprehensive statistical sample of variability data at sub-mm, IR, and X-ray wavelengths, the authors discussed a strongly variable one-zone synchrotron model\footnote{In this scenario, this highly variable component contributes to the sub-mm but cannot explain the observed sub-mm flux density levels entirely. A second electron population is required to explain the SED at radio to sub-mm wavelengths, and the observed sub-mm flux density is the result of the superposition of both components.} at sub-mm to NIR wavelengths that explains the X-ray emission by inverse Compton emission. More precisely, sub-mm synchrotron photons are up-scattered to the X-ray regime by the same electron population that is responsible for the synchrotron emission. This model was motivated by the two facts that: 1) a compact, self-absorbed synchrotron source has the conditions necessary for the scattering efficiency to be significant, and 2) the mechanism can explain the observed flux densities in the sub-mm, IR, and X-ray; the respective power spectral densities; and the cross-correlation properties between these bands.

One shortcoming of the analysis of Witzel et al. (2021) is its inability to explain the IR spectral indices $\alpha > -0.8$ as observed for several bright flares, among which is the flare discussed here. This is a consequence of relating the amplitude of the variable flux densities at IR and sub-mm wavelengths. In this model, the IR and sub-mm flux densities have been related to explain the strong correlation of X-ray photons (which are up-scattered from the sub-mm) with the IR. While this model was proposed as a baseline model that works for moderate flares at flux densities where the IR spectral indices are also described properly, the Witzel et al. (2021) speculated that brighter flares with blue spectral indices are states in which up-scattered photons contribute to the SED even in the IR.

We implemented an SSC model based on a non-thermal, power-law-distributed electron energy distribution to fit the time-resolved data of 2019 July 18. This model was determined by the same parameters as the PLCool$\gamma_{max}$ model, but it differs fundamentally from the synchrotron models above: the synchrotron part of the spectrum is located in the sub-mm (i.e., the SSC model predicts correlated sub-mm variability during this IR and X-ray flaring episode), and the IR and X-ray emission is explained through inverse Compton up-scattered photons. 

In this case the parameters are also degenerate: at different electron densities $n_e$ the source parameters $B$, $R$, and the energy range $\gamma_{min}$ to $\gamma_{max}$ can be chosen such that the IR to X-ray inverse Compton spectrum is reproduced as measured. However, for $n_e < 10^9~\mathrm{cm^{-3}}$ the synchrotron component will significantly exceed observed sub-mm emission levels. Therefore, we fixed the slope of the electron energy distribution to $p = 3.1$, which is consistent with the posterior of the analysis by Witzel et al. (2021). We then chose initial conditions with tight bounds such that: 1) The sub-mm luminosity remains within the range of observed sub-mm flares, and 2) all parameters show a continuous progression in time. 

\begin{figure}
    \centering
    \includegraphics[width=0.5\textwidth,angle=-0]{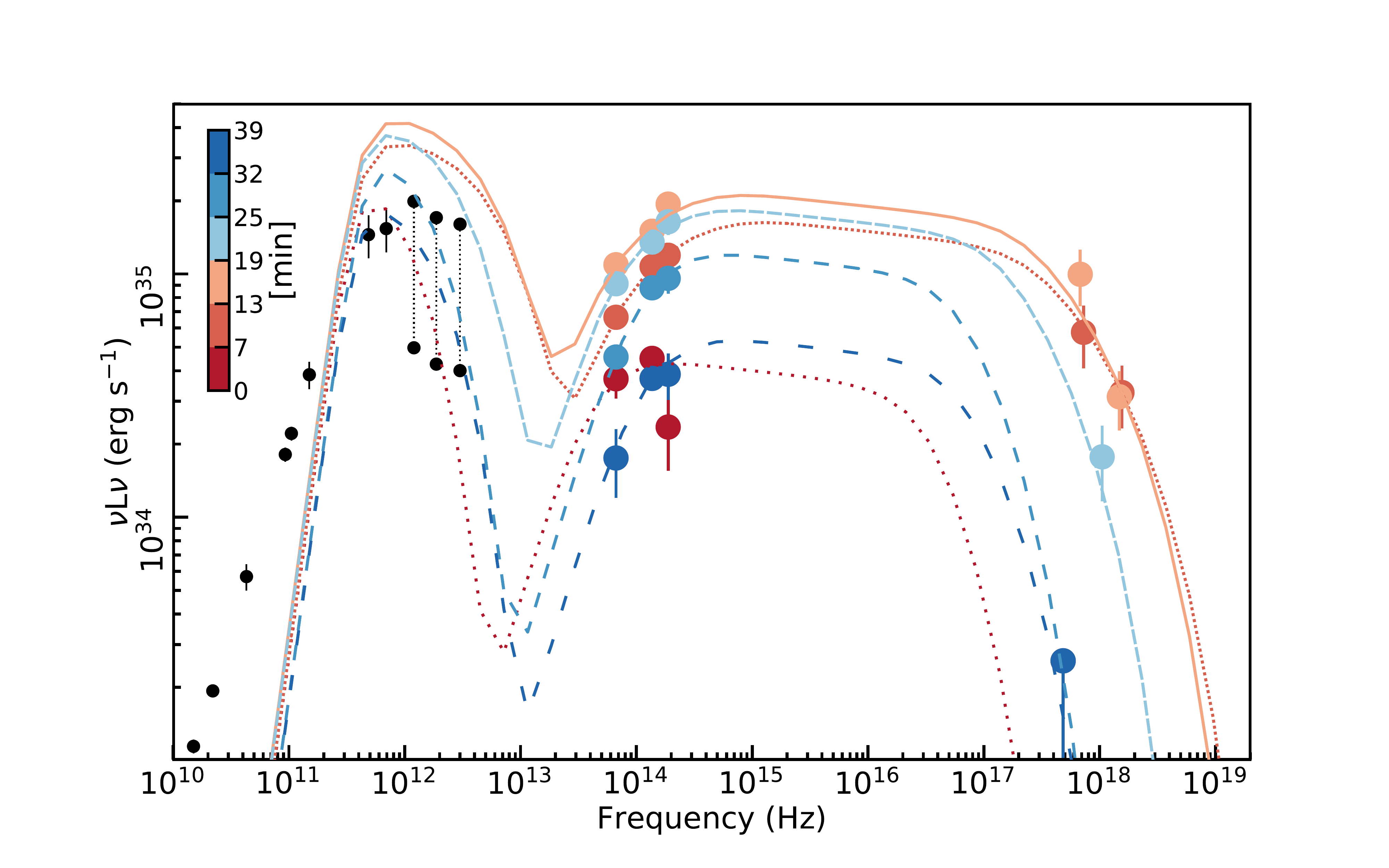}
    \caption{SEDs of the best fit synchrotron self-Compton models. Colors correspond to T1 to T6 as shown in the colorbar. The colored point show the observed data for each time. The dark points show the sub-mm SED of Sgr~A* with the same data as in Figure \ref{fig:SED}. As in the models involving only synchrotron emission, the flare evolution can largely be explained by progression of the electron density $n_e$ and high energy cutoff $\gamma_{\rm{max}}$.}
    \label{fig:TRSED_SSC}
\end{figure}

\begin{figure}
    \centering
    \includegraphics[width=0.5\textwidth,angle=-0]{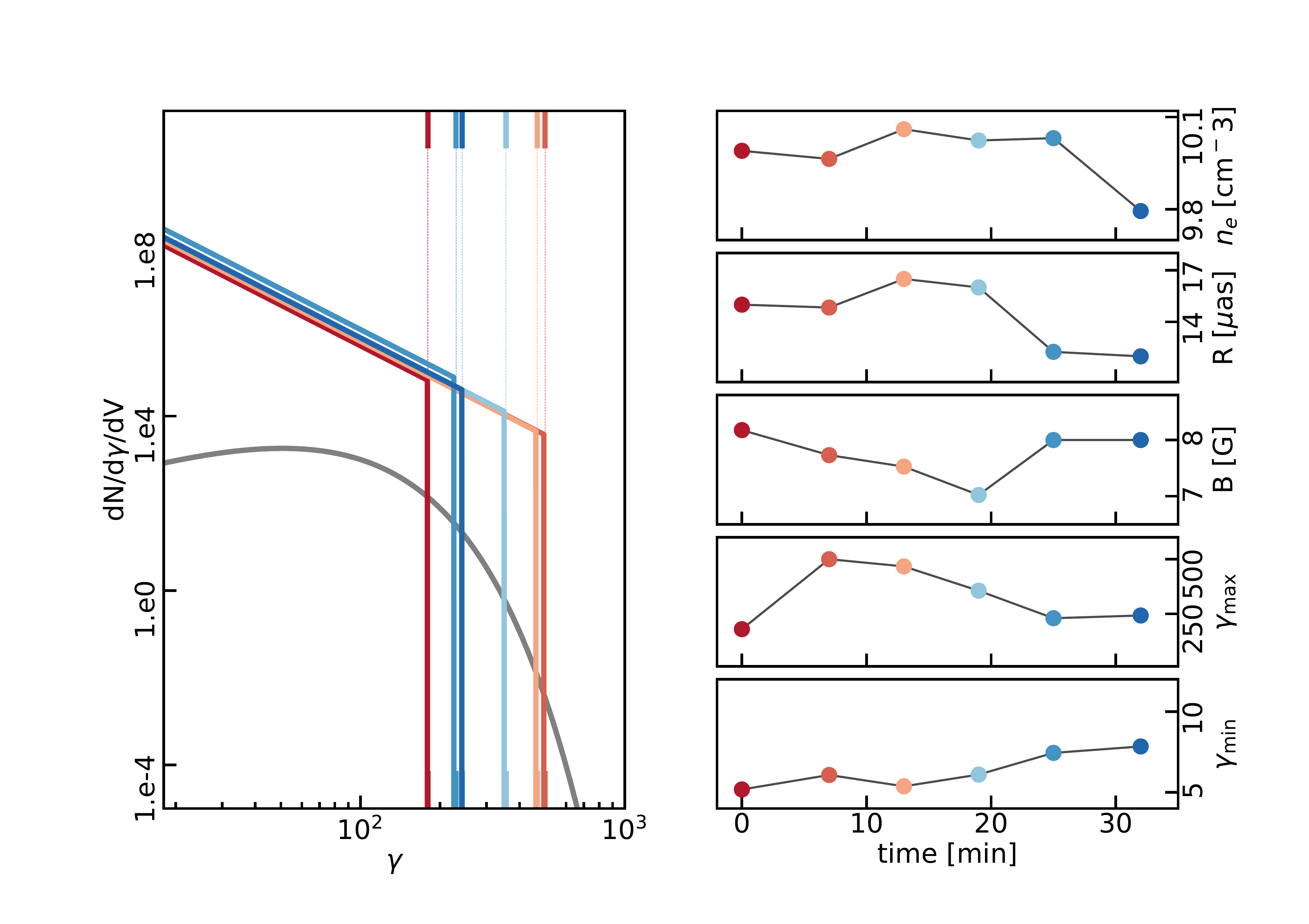}
    \caption{SSC Parameter evolution during flare, analogous to Figure~\ref{fig:electron_dist_1col}. Left: The evolution of the electron distribution during the flare. The dashed dotted lines indicate the the location of $\gamma_{max}$. The grey line shows the thermal distribution of electrons, peaking at $\gamma\sim 50$, which sets the minimum acceleration of the electrons for the flare. Right: evolution of the model parameters $n_{e}$, $R$, $B$, $\gamma_{max}$, and $\gamma_{min}$. For the SSC models $\gamma_{\rm{max}}$ is significantly lower and $n_e$ significantly higher than for the \plcg{} model.}
    \label{fig:TRSED_SSC_param}
\end{figure}

In T2--T4, where X-ray emission was detected, all other parameters besides $p$ were left free in the fits of the SEDs. We additionally fixed $\gamma_{max}$ in T1 and T5, $R$ for T1, T4, and T6, and $B$ for T5 and T6. To derive reliable uncertainties for T2--T4, we probed the parameter space with an MCMC sampler after lifting the bounds. The results are listed in Table~\ref{tab:SSC}, and the resulting SEDs and time series of parameters are shown in Figures~\ref{fig:TRSED_SSC} and \ref{fig:TRSED_SSC_param}.

\begin{table*}
\begin{center}
\begin{tabular}{ l c c c c c c }
\hline
\hline
 & \multicolumn{6}{c}{\bf Time Resolved} \\
 &  T1 & T2 & T3 & T4 & T5 & T6 \\
\hline
$log(n_{e}\times~1\mathrm{cm^{-3})})$& 10.0 & 10.0$\pm$ 0.5 & 10.1$\pm$0.4 & 10.0$\pm$0.2 & 10.0 & 9.8\\
$R~\mathrm{\mu as}$          & 15\dag & 15$\pm$8 & 16$\pm$8 & 16\dag & 12 & 12\dag\\     
$B~\mathrm{G}$               & 8.2 & 8$\pm$6 & 8$\pm$5 & 7$\pm$10 & 8.0\dag & 8.0\dag \\
$p$               & 3.1\dag & 3.1\dag & 3.1\dag & 3.1\dag & 3.1\dag & 3.1\dag\\
$\gamma_{max}$    & 180\dag & 500$\pm$100 & 470$\pm$80 & 360$\pm$70 & 230\dag & 243\\
$\gamma_{min}$    & 5.2 & 6.1$\pm$1.8 & 5.4$\pm$1.1 & 6.1$\pm$0.9 & 7.4 & 7.8\\
$\chi_{red}^2$; DOF& 6.7 & 0.1 & 2.7 & 0.9 & 0.2 & 0.4\\
\hline
\hline
\end{tabular} 
\caption{Best fit parameters of the fit of the SED with the SSC model. 
$n_{e}$: electron density within the source; 
$p$: power law index of the electron distribution; 
$R$: projected radius, in $\mathrm{\mu as}$, of the emitting source; 
$B$: magnetic field intensity (G); 
$\gamma_{max}$: maximum Lorentz factor of the accelerated electrons;
$\gamma_{min}$: minimum Lorentz factor of the accelerated electrons;
$\chi^2_{red}$; DOF: reduced $\chi^2$ and number of free parameters of the best fit
\dag: value fixed.
}
\label{tab:SSC}
\end{center}
\end{table*} 

\section{Discussion}
This paper discusses the first Sgr~A* flare that has been continuously observed from $4.5~\mathrm{\mu m}$ to $1.65~\mathrm{\mu m}$ in the near-infrared and from $2~\mathrm{keV}$ to $70~\mathrm{keV}$ in the X-ray band. Compared to previously studied flares simultaneously observed in the X-ray and IR bands, this flare is exceptional for its: i) remarkable IR brightness; ii) relative X-ray faintness, and iii) short duration. 

\subsection{Slope variability in the IR band during the flare}
The IR spectrum of the flare showed an increasing spectral index with increasing flux density. During the onset of the flare, the ratio of the $H$-band flux to the $M$- and $K$-band fluxes was low. This resulted in a kink in the intra-IR spectrum. The $H-K$ slope seemed to increase with flux density, being the bluest when the flare was the brightest and decreased again towards the end of the flare. Such a flux correlation has been discussed in previous works. While \cite{Hornstein2007} measured a constant spectral slope $\nu F_\nu\propto \nu^{0.5}$ independent of flux density, \cite{Eisenhauer2005}, \cite{Gillessen2006}, and \cite{Genzel2010} confirmed  $\nu F_\nu\propto \nu^{0.5}$ at high flux density but argued for a flux-dependent  $\nu F_\nu\propto \nu^{-1\dots -3}$ at lower flux density. The statistical analysis of the $M$- and $K$-band flux distributions presented in \cite{Witzel2018} favoured a variable, flux-dependent spectral index. Our work adds further evidence for a flux-dependent spectral index. Small changes in the spectral slope that be explained either by stochastic fluctuation or a flux-dependent scaling. We also found a kink in the intra-IR spectral slope during T1. Despite the difficulty of obtaining reliable flux measurements at very low flux from AO photometry, the variation is formally significant (${>}1~\sigma$, \autoref{fig:temporal SED}). 

\subsection{A single zone emission model for \sgras}

Using our fast numerical implementation of a one-zone emitting source, we explored a variety of models, at first regardless of their physicality in the context of Sgr~A*'s accretion flow. All our models require a set of parameters describing the ambient conditions: i) electron density $n_e$; ii) magnetic field strength $B$; iii) radius $R$ of the emitting source (assumed to be spherical), and iv) a energy distribution of accelerated electrons described by a set of parameters. 
For a thermal scenario, the distribution is characterized by a single parameter: the temperature of the electrons. For a power-law distribution, at least two parameters are required: the slope of the distribution and one or two normalisation constant ($\gamma_{min}$, $\gamma_{max}$). The normalisation constants can be interpreted in a physical sense: if the distribution is generated from a process which accelerates particles, then the minimum Lorentz factor $\gamma_{min}$ can be interpreted as the ambient Lorentz factor of the particles. 
Similarly, the maximum Lorentz factor $\gamma_{max}$ can be interpreted as a maximum length scale on which the particles are accelerated. 
Furthermore, the power-law distribution can have more than one slope. Such a broken power-law distribution is for instance assumed in the \plcg{} model, where synchrotron cooling is expected to induce a change of $p_2 = p_1 - 1$ at the cooling break. 

Before reaching the observer, the synchrotron radiation can be up-scattered by a population of relativistic electrons and produce an inverse Compton component. 
For example, for synchrotron one-zone models that take into account the respective synchrotron self-Compton component, there are three different ways of obtaining simultaneous IR and X-ray emission. 

\begin{enumerate}
    \item The emission in both bands is entirely dominated by synchrotron emission. We refer to scenarios of this type as SYN--SYN scenario.
    In such scenarios, the photon index observed in the X-ray band should be steeper by 0.5 than the simultaneous IR value (as a consequence of the cooling break).
    \item The emission in the IR is synchrotron emission, and the X-ray emission is synchrotron self-Compton emission. We refer to these scenarios as SYN--SSC scenario.
    \item The emission in both bands is entirely dominated by the inverse Compton component of the synchrotron self-Compton emission. We refer to such a scenario as SSC--SSC scenario.
\end{enumerate}

\subsection{Constraints from the simultaneous IR and X-ray photon indices and flux ratios}
The combination of: i) the observed positive IR slope; ii) the observed negative X-ray slope, and iii) the large flux ratio between IR and X-ray is a major problem for the SYN--SYN as well as the SSC models. 

Taken at face value, the difference in X-ray to IR slopes would be perfectly consistent with a synchrotron model with a cooling break in the electron distribution \citep{Dodds-Eden2009}. However, such a model cannot at the same time reproduce the flux ratio between the IR and X-ray (\S 8.1).

The observed luminosity in both bands sets parameter regimes for which the three scenarios match the observed spectrum:

\begin{itemize}
    \item In order to be dominated by synchrotron emission in both bands, the maximum Lorentz factor $\gamma_{max}$ is required to be $\gg10^4$.
    \item In order to be dominated by synchrotron emission in the IR and by SSC in the X-ray, $\gamma_{max}$ must be rather low. The frequency at which the synchrotron emission peaks scales $\nu_c (B) \times \gamma_{max}^2$. Therefore a large magnetic field $\gg10^3$~G is needed to shift the synchrotron peak into the IR.
    \item Similarly, in order to be dominated by SSC in both bands, $\gamma_{max}$ cannot be too large. However, because the synchrotron emission does not need to be shifted into the near-infrared, the constraints on the magnetic field can be relaxed. Nevertheless, in order to sustain high SSC flux from IR to X-ray, the particle density has to be $\gg10^9$~cm$^{-3}$. 
\end{itemize}

\subsubsection{The SYN--SSC scenario:}

The SYN--SSC scenario has severe problems: First, it requires magnetic fields of ${\sim} 10^4~\mathrm{G}$, source regions around ${\sim} 0.001 \mathrm{R_s}$, and densities ${\sim} 10^{12}~\mathrm{cm^{-3}}$. These parameters are extreme compared to the sub-mm ambient conditions. Even ignoring this, the synchrotron cooling time scale in such a strong magnetic field is on the order of $0.1$ seconds in the IR and of the order of $1$ millisecond in the X-ray. 
Despite flares of Sgr~A* being highly variable, spikes on time-scales shorter than tens of seconds have never been observed in the IR band. We attribute this lack of short time-scale IR variability to the effects of the cooling time of the electrons, which smooth out any variation shorter than a few seconds. 
We rule out, \cite{Dodds-Eden2009} and \cite{Dibi2014}, the scenario in which the IR flare is generated from synchrotron emission with a thermal distribution, and the X-ray flare is synchrotron self-Compton. This is a direct consequence of the negative X-ray spectral slope. If the observed X-ray slope were flat or positive, the requirement of a $\gamma_{max}<10^2$ would be relaxed. This is because for a positive or flat spectral slopes the emission can stem from the rising or flat part of SSC spectrum. In turn, this relaxes the requirement for very large magnetic fields, because the peak of the synchrotron component at $\nu_{max,syn}$ can be shifted by $\gamma_{max}$ as well and not only by the magnetic field.

\subsubsection{The SSC--SSC scenario:}

In the picture of the time-dependent model of Witzel et al. (2021), which can successfully describe the flux density distributions and the auto-correlation and cross-correlation properties of the light curves, the fast IR variability is the result of a quickly varying $\gamma_{max}$ that truncates the synchrotron spectrum. In order to link the IR variability amplitudes at longer timescales with the sub-mm and X-ray regimes, an overall $\alpha = -1$ is required that---depending on the brightness---steepens towards the IR due to the $\gamma_{max}$ cutoff. Flatter IR spectral indices of $\alpha > -0.8$ as reported here are not possible without the up-scattered spectrum contributing to the IR. 

The 2019-07-17 flare requires an even more extreme scenario in that it shows a very bright IR flare in combination with moderate X-ray luminosity. This particular configuration requires the range of the SSC component of the spectrum to be limited such that its decreasing flank falls into the 2--8 keV range. For the fit, this is achieved by restricting $\gamma_{max}$ to lower values such that the IR is not a superposition of direct synchrotron and scattered photons anymore but is dominated by the SSC component entirely.
To then reach the high IR flux density of this flare while keeping $B$ and $R$ at levels that do not lead to unobserved, high sub-mm luminosities, $n_e>10^{10}~\mathrm{cm^{-3}}$ is required. While much higher than the typical, average electron densities derived from modeling the radio to sub-mm SED of Sgr~A* with synchrotron emission from a thermal electron distribution (ambient $n_e<10^7~\mathrm{cm^{-3}}$, \citealt{Bower2019}), $n_e>10^{10}~\mathrm{cm^{-3}}$ is not out of the question: \cite{Moscibrodzka2013} discussed mid-plane densities of $n_e=10^9~\mathrm{cm^{-3}}$, and \cite{2020MNRAS.499.3178Y} used $10^{-13}~\mathrm{gcm^{-3}}$, which corresponds to $5.9\cdot10^{10}~\mathrm{cm^{-3}}$.

\subsubsection{The SYN--SYN scenario:}

The SYN--SYN scenario realized via the \plcg{} model requires $\gamma_{max}\sim 50~000$ and an exponential decay rather than a sharp cutoff (see \autoref{fig:TRSED_all}). Because our data constrains the fit in the optically thin part of the spectrum, we can infer only the total number of electrons rather than the radius and electron density independently. Fixing the source radius to $1R_S$, we obtained an estimate of the electron density. By assuming a cooling time scale of 2 minutes and by requiring a cooling break between the IR and the X-ray, the magnetic field is constrained to $B\sim1$ to $100~\mathrm{G}$\footnote{This is sensitive to our choice of the cooling timescale because the break frequency scales as $\nu_{break}\propto 1/t_{cool}^2$.}. Under these assumptions, the plasma parameters required are comparable to the sub-mm ambient parameters inferred from the sub-mm SED (e.g.\ \citealt{Yuan2003, Bower2019}). This model requires that the process accelerating the electrons: i) generate Lorentz factors increased from ambient conditions by a factor $>10^3$ ; ii) do so without alteration of the ambient plasma parameters on large scales. The best-fit model for the mean SED sets a direct constraint on $\gamma_{max}$. As discussed in section \ref{sec:plcg model}, this is a consequence of the high flux in the IR together with moderate flux in X-ray. 
Under the model assumptions, our observations place limits on the maximum acceleration of the flare-generating process (as done by \citealt{Ponti2017}). Notably, this flare mechanism does not produce any relevant sub-mm flux. Therefore, it does not predict any direct effect on the sub-mm light curve and observable accretion flow\footnote{This is strictly true only if the assumptions made here are valid. \cite{Ponti2017} discussed a brighter X-ray flare, where the magnetic field strength was consistent with the ambient value before and after the flare, while it significantly drops at the peak of the flare. If the magnetic field strength dropped at the peak of the flare (possibly as a consequence of magnetic re-connection) in a significant fraction of the volume producing the emission in the sub-mm band, then a drop in the sub-mm emission might be expected to be observed at the peak of the flare as a consequence of the smaller magnetic field strength \citep{Dodds-Eden2010}.}. 
For our choice of $R = 1~R_s$, the SSC component of the flare peaks at around $10^{23}~\mathrm{Hz}$ (corresponding to GeV energy band), with a peak luminosity of $\sim10^{34}~\mathrm{erg~s^{-1}}$ (\autoref{fig:MeanSEDPLSSC}). Unfortunately, this implies that the expected SSC luminosity is too faint to be observable by, for instance, the \fermi{} satellite \citep{Malyshev2015}.

\subsection{The temporal evolution of the flare}

\subsubsection{Temporal evolution in the SSC--SSC scenario}

The Compton component of the SSC--SSC model is sensitive to where the synchrotron emission becomes optically thick. Therefore, such a model places strong constraints on the synchrotron part of the spectrum, which is expected to reproduce the emission in the sub-mm band. Unfortunately, our campaign has no coverage of the sub-mm band. Therefore, we cannot uniquely derive the best-fit solution but instead can only constrain the parameters by assuming typical values for the sub-mm emission. Keeping the magnetic field, the electron density and the radius thus constrained, we modeled the light curve of the flare by selecting a suitable local minimum. The temporal evolution of flux densities is then mostly driven by the variation of $\gamma_{max}$, which determines the width of the synchrotron spectrum and, as a consequence, scales the X-ray flux. 

The SSC--SSC scenario predicts that a high sub-mm flux density excursion is associated with the flare of 2019-07-17, i.e., that the sub-mm exhibits temporal correlation with the IR light curve. Depending on the exact combination of parameters, the sub-mm light curve may lag slightly behind the IR and X-ray, comparable to the effects of source expansion discussed by Witzel et al. (2021).

The `kink' in the X-ray spectrum of the first data point cannot be explained by SSC-SSC scenario because it either requires a too-narrow SSC component, or an extension of the synchrotron component into the IR for only the first data point. Except for this cutoff between the $K$- and $H$-band of T1, the model can closely fit the measurements. In particular, it reproduces the frequency-dependent spectral index in the IR that changes from the very blue index between the $M$- and $K$-band to a flatter $K-H$ index.

\cite{Bower2018} showed in a study of ALMA polarization data that the observed Faraday rotation is consistent with the rotation measure expected from a radiatively inefficient accretion flow (RIAF) with $\rm{\dot{M}} = 10^{-8} \rm{M}_{\odot} \rm{y}^{-1}$, or $\rm{\dot{M}} = 3 \cdot 10^{-16} \rm{M}_{\odot} \rm{s}^{-1}$. Assuming a proton to electron ratio of unity, the changes in electron density as suggested by the temporal evolution described here of $\Delta n_e \approx 6.3 \cdot 10^9~\mathrm{cm^{-3}}$ over a region of ${\sim}1.5\; R_S$ require an additional mass $\Delta M\approx 1.3 \cdot 10^{-10}\; \rm{M}_{\odot}$. The average accretion flow would require ${>}100$~hours to provide this much mass, but in this scenario the density evolves within less than 30 min. This suggests that interpreting the flare in the context of the SSC--SSC model makes the implicit assumption of moments of extraordinary accretion far exceeding the average accretion flow.  

\subsubsection{Temporal evolution of the SYN--SYN scenario}
The time-resolved spectra have been fitted assuming a constant magnetic field strength and source size because of the degeneracy with the electron density. Therefore, in our modeling, the normalisation of the spectrum is mainly determined by the electron density. Similar to the model discussed by \cite{Dodds-Eden2010} and \cite{Ponti2017}, this scenario assumes an episode of particle injection with large $\gamma_{max}$ which sustains the X-ray emission against the very short cooling time scales. The quality of the data and the degeneracy of the model parameters do not allow us to explicitly model the evolution of the radius and magnetic field intensity in addition to the electron density (e.g.\ \citealt{Dodds-Eden2010}; \citealt{Ponti2017}). Therefore, it remains to be verified whether the findings of \cite{Dodds-Eden2010} and \cite{Ponti2017} hold and are applicable here as well. 

Although it appears sharper than the model predicts, the apparent `kink' in the IR spectrum at T1 is attributed to the truncated electron distribution function at $\gamma_{max}\sim 500$. 
These observations place strong constraints on the time scales under which electron acceleration has to be maintained and on how fast it needs to vary (see \autoref{fig:electron_dist_1col}; \citealt{Ponti2017}). 

\subsection{Concluding remarks}

For both the SYN--SYN and SSC--SSC models, this flare sets strong requirements on the mechanism responsible for its emission: either it requires acceleration of electrons by a factor of ${>}10^3$, or it requires electron densities increased by a factor of $10^{2\dots3}~\mathrm{ecm^{-3}}$ and electron density changes with respect to the sub-mm ambient conditions that cannot be explained from the average accretion flow. Furthermore, it is remarkable that in both cases, the maximum Lorentz factor plays a very important role for the temporal evolution of the flare. For the SSC--SSC scenario, $\gamma_{max}$ regulates the width of the synchrotron spectrum, which in turn sets the width of the Compton component. Similarly, for the SYN--SYN scenario, the kink of the IR spectrum for T1, the high IR-to-X-ray flux ratio, and the X-ray slope are dictated by the evolution of $\gamma_{max}$. A similar evolution of the SED was observed during another flare detected simultaneously in the IR and X-ray band \citep{Ponti2017}. 
Both models make strong predictions about the presence of a direct sub-mm counterpart: the SSC--SSC scenario would be ruled out in the absence of a strong flux increase by a factor of two to three, while the extrapolation to the sub-mm band of the SYN--SYN model predicts no significant contribution to the sub-mm emission (although a possible variation of the magnetic field might induce some degree of correlated variations in the sub-mm band; \citealt{Dodds-Eden2010, Ponti2017}). All of our modeling has ignored the expected modulation of the light curve from the relativistic motion of the flare itself and other relativistic effects expected in the proximity of the black hole \citep{GravityCollaboration2018_orbital, GravityCollaboration2020_orbital}. For the SYN--SYN scenario, the modulation of the light curve by relativistic boosting will merely translate into a variation of the assumed parameters. The same is not true for the SSC--SSC scenario: the Compton scattering occurs in the flare rest frame, while the synchrotron emission is observed from outside. Consequently, the SSC component of a relativistic hot spot will be lowered while the synchrotron component may be increased (or vice versa). Future modeling of such a scenario should take this effect into account. 

In light of the new data, we rule out the SYN--SSC scenario for this flare because it requires nonphysical model parameters and would imply NIR variability on timescales not observed. We consider that neither the SYN--SYN nor the SSC--SSC models can be strictly ruled out. However, the SSC--SSC scenarios requires very high local over-densities in the accretion flow and a density variation that cannot be explained with the average mass accretion. It therefore requires an extraordinary accretion event together with moderate particle acceleration. 

The SYN--SYN model does not require extraordinary accretion, but requires particle acceleration from Lorentz factors of the ambient electrons of $\gamma\sim 10$ to $\gamma\sim 10^{4}$. Typically discussed candidate mechanisms are either electron acceleration through magnetic reconnection, turbulent heating, in shocks induced by a misalignment of black hole spin and accretion flow, or in shocks along an outflow/jet \citep{Dodds-Eden2009, Dexter2012}. Large scale simulations of the accretion flow do not have the resolution to trace individual reconnection events, but several strategies have been developed to try to account for this \citep{Dexter2020_model, Chatterjee2020}. 
Particle in cell simulations of plasmas show that both turbulence heating and magnetic reconnection can create significantly nonethermal, power-law electron distributions \citep{Sironi2020, Wong2020, Werner2021}.
Interestingly, the large scale simulation presented by \cite{Ripperda2020} shows flare regions of a size of around $1$ to $2R_S$ formed through magnetic reconnection with comparable field strengths to those in the toy models discussed here. In the SYN--SYN model, this flare places tight constraints on the maximum allowed acceleration. If no rigorous theoretical motivation for such a specific value of the maximum acceleration value is found\footnote{For instance, assuming the particles are accelerated for $1R_S$ with a fraction of the speed of light would yield a Lorentz factor of $\gamma(v) = eBR_S/m_ec^2 \times v/c \sim 1 \times 10^{10} \times v/c$, implying that the acceleration happens on much smaller scales.}, it may ultimately be viewed as too constraining to uphold the simple SYN--SYN model and it may need to be discarded in favour of more complicated models. Conversely, if the maximum acceleration of an acceleration process is rooted in a sound theoretical framework, future observations of infra-red bright and X-ray faint flares may provide a powerful tool to constrain the underlying acceleration physics. 

Currently, there are no models that can correctly match the observed spectrum, variability, and orbital motions of the emission at the Galactic Center. Our two models shown above reproduce the SED during flares, but do not include enough physics to account for variability or orbital motions. More physically motivated GRMHD simulations show more complexity but are also not able to fully explain observations. However, in GRMHD models the NIR synchrotron photons and inverse Compton scattering are associated with spatially separate populations of electrons, an effect that is not captured in our simple one-zone models. 
More work is needed to combine these approaches or develop new methods to understand the emission mechanism and dynamical properties of the accretion flow at the smallest scales.


\begin{acknowledgements}
SvF thanks Giulia Focchi for her contribution to the $H$-band acquisition camera pipeline. SvF, \& FW acknowledge support by the Max Planck International Research School. GP is supported by the H2020 ERC Consolidator Grant Hot Milk under grant agreement Nr. 865637. A.A. and P.G. were supported by Funda\c{c}\~{a}o para a Ci\^{e}ncia e a Tecnologia, with grants reference UIDB/00099/2020 and SFRH/BSAB/142940/2018. This work is based in part on observations made with the Spitzer Space Telescope, which was operated by the Jet Propulsion Laboratory, California Institute of Technology under a contract with NASA. Support for this work was provided by NASA.  The scientific results reported in this article are based in part on observations made by the Chandra X-ray Observatory. This work is based in part on observations made with NuSTAR, which is operated by NASA/JPL-Caltech. GGF, JLH, HAS, \& SPW acknowledge support for this work from the NASA ADAP program under NASA grant 80NSSC18K0416.
\end{acknowledgements}

\appendix
\section{Synchrotron self-Compton of the SYN--SYN scenario}
The synchrotron self-Compton component of the SYN--SYN scenario peaks at frequencies higher than the X-ray band. Unfortunately, for the parameter ranges we have assumed (Table \ref{tab:PLCoolgamma}), this peak is not bright enough to be detectable in the GeV bands by for example \fermi{} (e.g. \citealt{Malyshev2015}). However, it poses a possibility to constrain the radius and the particle density of the otherwise optically thin spectrum. At small enough radii and high enough densities, the falling flank of the SSC spectrum will start to contribute to the $2--70~\mathrm{keV}$ band of \nustar{}. For instance, at $B=30~\mathrm{G}$, the emission region is constrained to ${\sim} 0.3~\mathrm{R_S}$. This demonstrates the importance of further parallel NIR--X-ray observations with as wide as possible spectral range.
\begin{figure}
\centering
\includegraphics[width=0.45\textwidth,angle=-0]{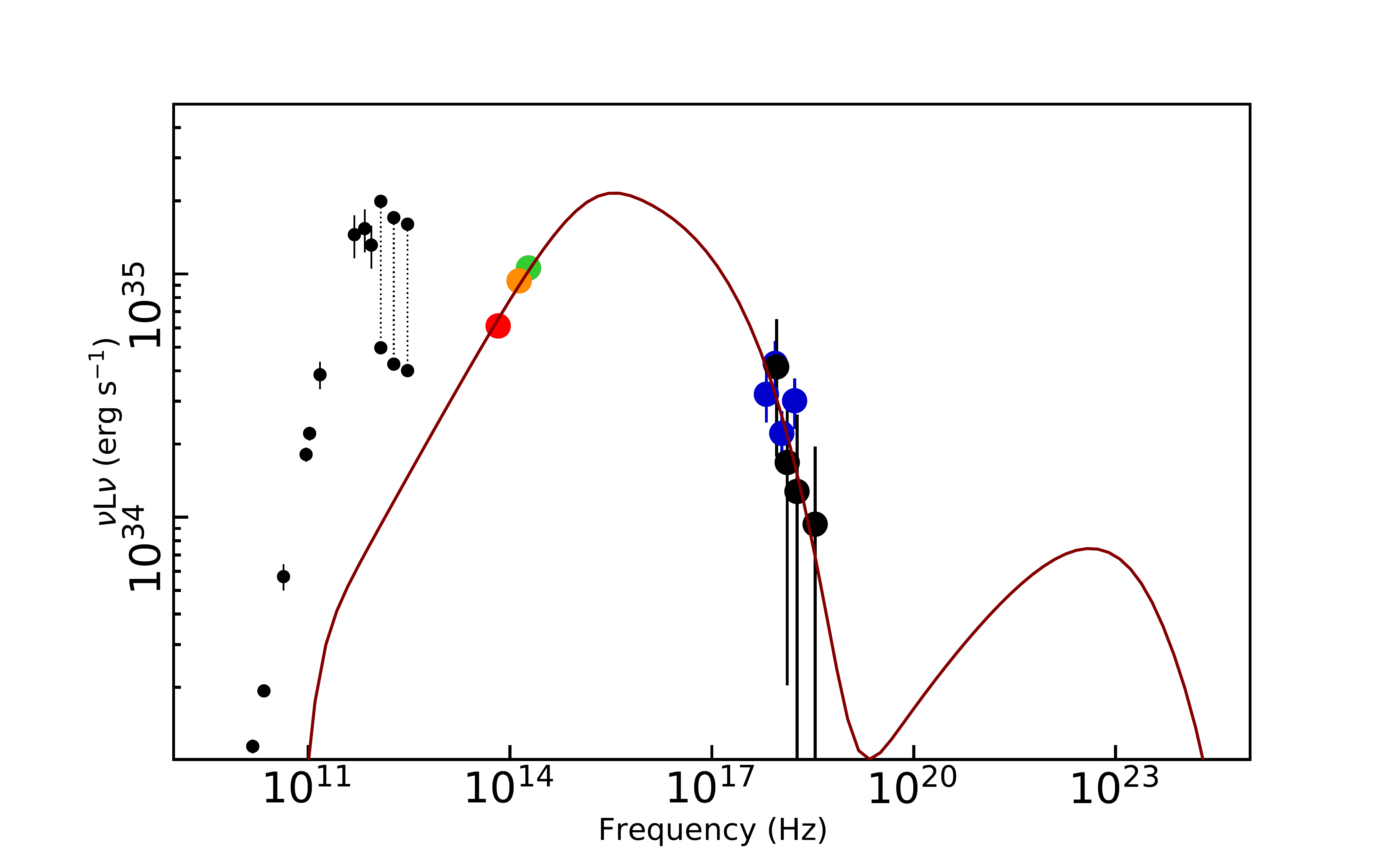}
\caption{Mean SED of Sgr~A* during the flare, including the synchrotron self-Compton component of the \plcg{} model. This component peaks at $\nu\sim 10^{23}~\mathrm{Hz}$.}
\label{fig:MeanSEDPLSSC}
\end{figure}

\section{Accounting for the acquisition camera transmission curve and the different spectral slopes of S2 and Sgr~A*}
\label{appendix:hband flux}
The GRAVITY flux measurements derived in both bands are measurements of the flux ratios of the S2 and \sgr{}. The spectral dependence of the reddened flux of Sgr~A* and S2 can be approximated as power-law with different indices:

\begin{equation}
\begin{aligned}
    &F_{S2}(\lambda) = F_{S2_0} \cdot \dfrac{\lambda}{\lambda_0}^{\alpha_{S2}}\\
    &F_{Sg}(\lambda) = F_{Sg_0} \cdot  \dfrac{\lambda}{\lambda_0}^{\alpha_{Sg}},
    \label{eqn:flux power law}
\end{aligned}
\end{equation}
where $F_{x_0}$ denotes the flux of the respective source at wavelength $\lambda_0$. For S2, the spectral slope $\alpha_{S2}$ can be determined from the NACO photometry in $H$- and $K$-band (e.g.\ \citealt{Gillessen2016}). 

To account for the effect of different spectral slopes on the flux ratio in $H$-band, we have to take the filter curves of the acquisition camera, the VLTI, and GRAVITY into account. This can be achieved by expressing the flux of both sources on the acquisition camera detector as function of the respective effective wavelengths:
\begin{equation}
\begin{aligned}
    &F_{S2}(\lambda) = F_{K, S2} \cdot \dfrac{\lambda_{eff, S2}}{\lambda}^{\alpha_{S2}}\\
    &F_{SgrA}(\lambda) = F_{K, SgrA} \cdot  \dfrac{\lambda_{eff, S2}}{\lambda}^{\alpha_{SgrA}}.
    \label{eqn:flux power law}
\end{aligned}
\end{equation}

Here the effective wavelength, assuming a power-law flux dependence, is given by:
\begin{equation}
\begin{aligned}
    \lambda_{eff}(\alpha) = \dfrac{\int F_\lambda(\alpha) \cdot \lambda ~ d \lambda}{\int F_\lambda(\alpha) d \lambda},
\end{aligned}
\end{equation}
where $F_\lambda = F_{source} (\alpha) \cdot T(\lambda)$ is the power-law source flux multiplied by the instrument transmission $T(\lambda)$. The observed flux ratio in the $H$-band can then be expressed as:
\begin{equation}
\begin{aligned}
     r_H = \dfrac{\int F_{K, S2} \cdot \left(\dfrac{\lambda_{eff, S2}}{\lambda_K}\right)^{\alpha_{S2}} d\lambda}{\int F_{K, SgrA} \cdot \left(\dfrac{\lambda_{eff, SgrA}}{\lambda_K}\right)^{\alpha_{SgrA}} d\lambda},
     \label{eqn:flux ratio relation}
\end{aligned}
\end{equation}
where $F_{K/H, S2}$ is the observed flux in $K$-band, and $\lambda_{eff, S2/SgrA}$ are the acquisition camera effective wavelength of S2 and Sgr~A*. We obtain $\lambda_{eff, S2}$ in the $H$-band using the acquisition camera transmission curve and the reddened power law flux relation determined from NACO photometry. 

Using the functional relation for the effective Sgr~A* wavelength in the $H$-band, we can rewrite this as:
\begin{equation}
\begin{aligned}
       \left(\dfrac{\lambda_{eff, SgrA}(\alpha_{SgrA})}{\lambda_K}\right) =  \left(\dfrac{\lambda_{eff, S2} (\alpha)}{\lambda_K}\right)^{\alpha_{S2}} \cdot \dfrac{r_K}{r_H},
\end{aligned}
\end{equation}
where $r_K$ and $r_H$ are the observed flux ratios in $H$- and $K$-band. We can numerically solve this equation for the effective wavelength $\lambda_{eff, SgrA}$.  Once $\lambda_{eff, SgrA}$ and $\alpha_{SgrA}$ have been determined, we can plug these into equation \ref{eqn:flux power law} to obtain the reddened $H$-band flux density $F_\lambda$. We converted $F_\lambda$ to flux density $F_\nu$ and deredden through the standard approach $F_{dered.} = F_{red.} \cdot 10^{0.4\cdot m_H}$, with $m_H$ as discussed in section \ref{sec:extinction}.

\section{Effect of the column density on the IR and X-ray spectral slope, and inferred parameters}\label{appendix:nh_density}
\begin{figure}
    \centering 
    \includegraphics[width=0.485\textwidth]{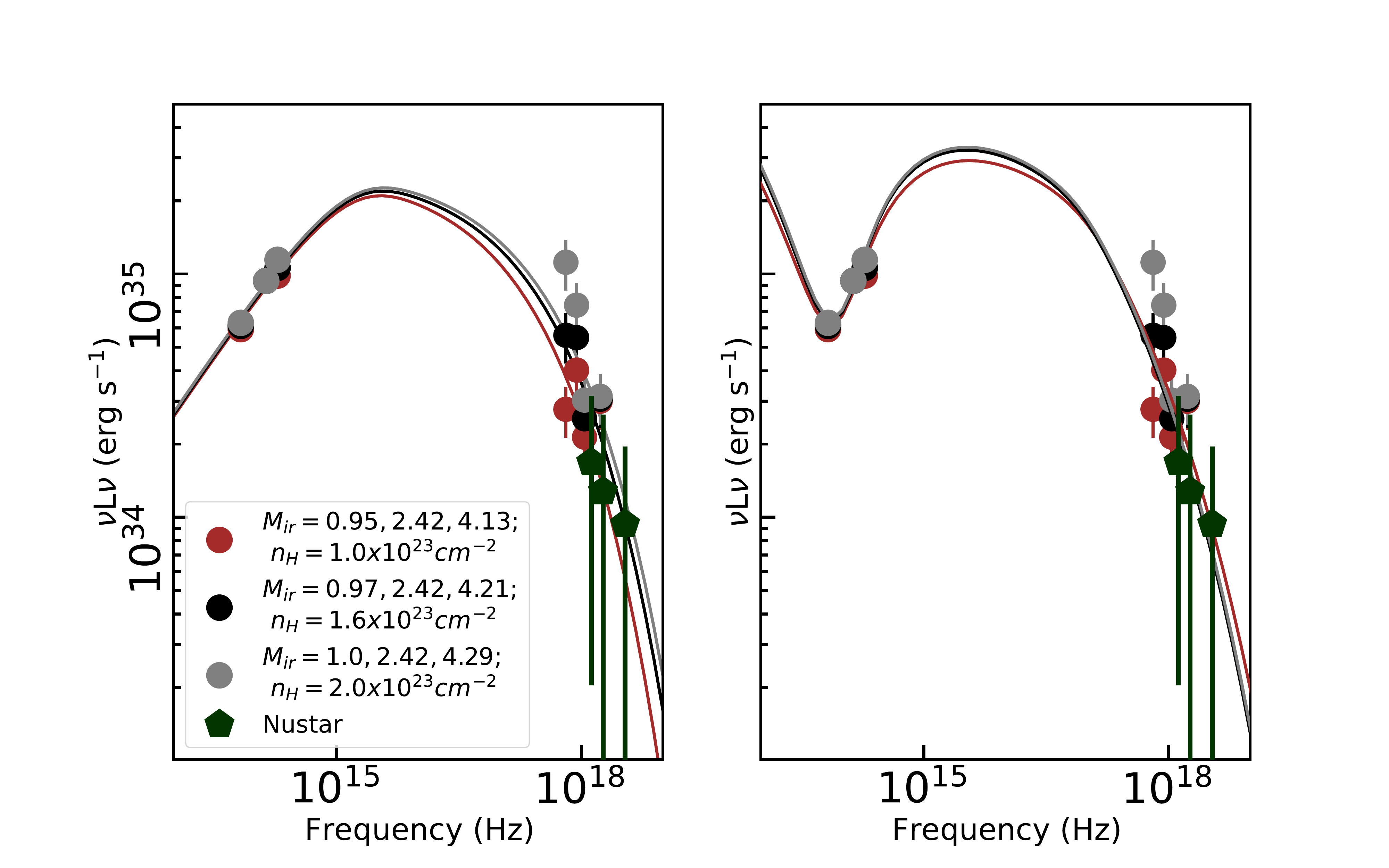}
    \caption{Effect of different neutral material column density: NIR data same as Figure \ref{fig:SED}. Both panels: the red, black and grey lines show the data corrected with different plausible neutral material column densities, which are reported in the legend of the left panel. The \nustar{} data have not been re-reduced (green pentagons), because the high energy data is only marginally affected. We've removed the lowest energy bin from the \nustar{} spectrum because it might be affected by the extinction. The models in the left panel are \plcg{} type models, on the right SCC-SSC type models are plotted, the color indicates the respective data set which is fitted.}
    \label{fig:absorption_nh_denstiy}
\end{figure}

As discussed in Section \ref{sec:chandra}, we chose three different column density values $n_{\rm{H}} = 1.0 \times 10^{23}\mathrm{cm^{-2}}$, $n_{\rm{H}} = 1.6 \times 10^{23}\mathrm{cm^{-2}}$, $n_{\rm{H}} = 2.0 \times 10^{23}\mathrm{cm^{-2}}$. We fitted the \chandra{} mean spectrum of the flare assuming each of the above mentioned values of the column density and computed the respective corrections in order to de-absorb the spectrum. Similarly, we vary the infrared extinction and scale the flux density according to the uncertainties reported in Table \ref{tab:extinction}. Figure \ref{fig:absorption_nh_denstiy}, shows the de-absorbed data and the resulting fits to the data sets. For the SYN--SYN model, we assumed the same parameters as for the \plcg{} model (see Table \ref{tab:PLCoolgamma}). For the SSC--SSC model, we assumed fixed radius $1R_S$, a fixed slope of the electron distribution $p=3.1$, and a fixed minimum acceleration $\gamma_{min}=10$. We fit the particle density, the magnetic field, and the maximum acceleration. Table \ref{tab:absoprtion_effect} reports the best-fit results. While the inferred parameters of the best-fit solution change slightly, the main conclusions of the paper are not affected by the choice of the specific extinction value: the SYN--SYN model requires a $\gamma_{max}\sim 10^4$ in order to explain the observed flux ratios in the NIR and X-ray. In contrast, in one-zone models, the SSC--SSC scenario requires particle densities $10^3$ higher than typically inferred for the ambient accretion flow. 

\begin{table*}
    \centering
    \begin{tabular}{c||c||c}
          & SYN--SYN model & SSC--SSC model\\
          Parameter & $\log_{10}(n_e)$ $\gamma_{max}$ & $\log_{10}(n_e)$, $B$, $\gamma_{max}$\\
         \hline
         $M_{IR}=0.95, 2.42, 4.13;~n_H=1.0\times 10^{23}~\mathrm{cm^{-2}}$&$6.243 \pm 0.015$, $39620 \pm 3808$&$9.75\pm0.03$, $(17.1\pm 0.25)~\mathrm{G}$, $276\pm29$\\
         $M_{IR}=0.97, 2.42, 4.21;~n_H=1.6\times 10^{23}~\mathrm{cm^{-2}}$&$6.240 \pm 0.011$, $47179 \pm 3824$&$9.74\pm0.02$, $(19.2\pm 0.3)~\mathrm{G}$, $244\pm13$ \\
         $M_{IR}=1.0, 2.42, 4.29;~n_H=2.0\times 10^{23}~\mathrm{cm^{-2}}$&$6.249 \pm 0.014$, $51113 \pm 5436$&$9.74\pm0.01$, $(19.5\pm 0.1)~\mathrm{G}$, $245\pm14$ \\
    \end{tabular}
    \caption{Effect of different choices of the neutral absorption column density. Fit parameters for SYN--SYN model and SSC--SSC model derived from least squares fitting. The models are described in Appendix \ref{appendix:nh_density}. The reported uncertainties were derived from the covariance matrix. }
    \label{tab:absoprtion_effect}
\end{table*}

\section{Analytical formulation of the non-thermal electron distributions}

We considered non-thermal electron distributions for the modeling of the flare SED in this paper. These are either in the form of a plain power-law or a broken power-law. In this section, we describe the analytical form of these distributions.

\begin{itemize}
    \item The formulation of the \textit{power-law} electron distribution is given in Equation \ref{eq:edist_pl}. In that equation, $n_e$ is the electron density, $p$ is the power-law index, and $\gamma_{min}$ and $\gamma_{max}$ are the low and high energy limits of the electron population. 
    \begin{equation} \label{eq:edist_pl}
    \frac{d n_{\text{pl}}}{d\gamma}= n_e \times
    \begin{cases}
        N_{\text{pl}} \gamma^{-p}, & \text{if } \gamma_{min} \leq \gamma \leq \gamma_{max}\\
        0, & \text{otherwise},
    \end{cases}
    \end{equation}

    \noindent where $N_{\text{pl}}$ is the normalization of the distribution,

    \begin{equation}
        N_{\text{pl}} = \frac{p-1}{{\gamma_{min}}^{1-p}-{\gamma_{max}}^{1-p}}.
    \end{equation}
    
    \item We provide a generic formulation of a \textit{broken power-law} electron distribution in Equation \ref{eq:edist_bpl}. Since we consider synchrotron cooling as the origin of the break at $\gamma_b$, we enforced $p_2=p_1+1$. For readability, we have used the notation $p$ in the main text for all power-law indices. In the case of cooled synchrotron spectra it corresponds to $p_1$.
    \begin{equation} \label{eq:edist_bpl}
    \frac{d n_{\text{bpl}}}{d\gamma}= n_e \times
    \begin{cases}
        N_{\text{bpl}} \gamma^{-p_1}, & \text{if } \gamma_{min} \leq \gamma\leq \gamma_b\\
        N_{\text{bpl}} \gamma^{-p_2} \gamma_b^{p_2-p_1}, & \text{if } \gamma_{b} < \gamma \leq \gamma_{max} \\
        0, & \text{otherwise},
    \end{cases}
    \end{equation}

    \noindent where $N_{\text{bpl}}$ is the normalization of the distribution,

    \begin{equation}
        N_{\text{bpl}} = \left[\left(\frac{\gamma_{min}^{1-p_1} - \gamma_{b}^{1-p_1}}{p_1-1}\right) + (\gamma_b)^{p_2-p_1}\left(\frac{\gamma_{b}^{1-p_2} - \gamma_{max}^{1-p_2}}{p_2-1}\right)\right]^{-1}
    \end{equation}
\end{itemize}

Considering synchrotron cooling in the presence of particle escape as the origin for the broken power-law distribution, a sharp cooling break in the electron distribution is not physical. However, the exact determination of the spectral shape is beyond the scope of this work. Furthermore, our observational data does not provide useful constraints on the cooling break itself, and thus the determination of the proper shape of the break is not required. For simplicity, we use the form given in Equation \ref{eq:edist_bpl}.

In above formulas, the electron distributions are truncated at both $\gamma_{min}$ and $\gamma_{max}$. As a more physical alternative, we use the an \textit{exponential cutoff} instead of a sharp truncation at $\gamma_{max}$,

\begin{equation}
    \frac{d n_{\text{expc}}}{d\gamma} = 
    \begin{cases}
        (d n/d\gamma) \exp(-\gamma/\gamma_{max}),  & \text{if } \gamma_{min} \leq \gamma \leq 10\times\gamma_{max}\\
        0, & \text{otherwise},
    \end{cases}
\end{equation}

\noindent i.e., we simply smooth the high energy cutoff of the original electron distributions with an exponential function. The high energy limit is extended from $\gamma_{max}$ to $10\times\gamma_{max}$.

\bibliography{theXrayFlare.bib}

\begin{thebibliography}{88}
\expandafter\ifx\csname natexlab\endcsname\relax\def\natexlab#1{#1}\fi

\bibitem[{{Baganoff} {et~al.}(2001){Baganoff}, {Bautz}, {Brandt}, {Chartas},
  {Feigelson}, {Garmire}, {Maeda}, {Morris}, {Ricker}, {Townsley}, \&
  {Walter}}]{Baganoff2001}
{Baganoff}, F.~K., {Bautz}, M.~W., {Brandt}, W.~N., {et~al.} 2001, \nat, 413,
  45

\bibitem[{Baganoff {et~al.}(2003)Baganoff, Maeda, Morris, Bautz, Brandt, Cui,
  Doty, Feigelson, Garmire, Pravdo, Ricker, \& Townsley}]{Baganoff2003}
Baganoff, F.~K., Maeda, Y., Morris, M., {et~al.} 2003, The Astrophysical
  Journal, 591, 891

\bibitem[{Barri{\`{e}}re {et~al.}(2014)Barri{\`{e}}re, Tomsick, Baganoff,
  Boggs, Christensen, Craig, Dexter, Grefenstette, Hailey, Harrison, Madsen,
  Mori, Stern, Zhang, Zhang, \& Zoglauer}]{Barriere2014}
Barri{\`{e}}re, N.~M., Tomsick, J.~A., Baganoff, F.~K., {et~al.} 2014, The
  Astrophysical Journal, 786, 46

\bibitem[{{Bouffard} {et~al.}(2019){Bouffard}, {Haggard}, {Nowak}, {Neilsen},
  {Markoff}, \& {Baganoff}}]{Bouffard2019}
{Bouffard}, {\'E}., {Haggard}, D., {Nowak}, M.~A., {et~al.} 2019, \apj, 884,
  148

\bibitem[{Bower {et~al.}(2018)Bower, Broderick, Dexter, Doeleman, Falcke, Fish,
  Johnson, Marrone, Moran, Moscibrodzka, Peck, Plambeck, \& Rao}]{Bower2018}
Bower, G.~C., Broderick, A., Dexter, J., {et~al.} 2018, The Astrophysical
  Journal, 868, 101

\bibitem[{Bower {et~al.}(2019)Bower, Dexter, Asada, Brinkerink, Falcke, Ho,
  Inoue, Markoff, Marrone, Matsushita, Moscibrodzka, Nakamura, Peck, \&
  Rao}]{Bower2019}
Bower, G.~C., Dexter, J., Asada, K., {et~al.} 2019, The Astrophysical Journal,
  881, L2

\bibitem[{Bower {et~al.}(2015)Bower, Markoff, Dexter, Gurwell, Moran,
  Brunthaler, Falcke, Fragile, Maitra, Marrone, Peck, Rushton, \&
  Wright}]{Bower2015}
Bower, G.~C., Markoff, S., Dexter, J., {et~al.} 2015, The Astrophysical
  Journal, 802, 69

\bibitem[{{Boyce} {et~al.}(2019){Boyce}, {Haggard}, {Witzel}, {Willner},
  {Neilsen}, {Hora}, {Markoff}, {Ponti}, {Baganoff}, {Becklin}, {Fazio},
  {Lowrance}, {Morris}, \& {Smith}}]{Boyce2019}
{Boyce}, H., {Haggard}, D., {Witzel}, G., {et~al.} 2019, \apj, 871, 161

\bibitem[{Brinkerink {et~al.}(2015)Brinkerink, Falcke, Law, Barkats, Bower,
  Brunthaler, Gammie, {Violette Impellizzeri}, Markoff, Menten, Moscibrodzka,
  Peck, Rushton, Schaaf, \& Wright}]{Brinkerink2015}
Brinkerink, C.~D., Falcke, H., Law, C.~J., {et~al.} 2015, Astronomy and
  Astrophysics, 576, 41

\bibitem[{{Chatterjee} {et~al.}(2020){Chatterjee}, {Markoff}, {Neilsen},
  {Younsi}, {Witzel}, {Tchekhovskoy}, {Yoon}, {Ingram}, {van der Klis},
  {Boyce}, {Do}, {Haggard}, \& {Nowak}}]{Chatterjee2020}
{Chatterjee}, K., {Markoff}, S., {Neilsen}, J., {et~al.} 2020, arXiv e-prints,
  arXiv:2011.08904

\bibitem[{Dexter \& Fragile(2012)}]{Dexter2012}
Dexter, J. \& Fragile, P.~C. 2012, Mon. Not. R. Astron. Soc, 000, 1

\bibitem[{{Dexter} {et~al.}(2020){Dexter}, {Jim{\'e}nez-Rosales}, {Ressler},
  {Tchekhovskoy}, {Baub{\"o}ck}, {de Zeeuw}, {Eisenhauer}, {von Fellenberg},
  {Gao}, {Genzel}, {Gillessen}, {Habibi}, {Ott}, {Stadler}, {Straub}, \&
  {Widmann}}]{Dexter2020_model}
{Dexter}, J., {Jim{\'e}nez-Rosales}, A., {Ressler}, S.~M., {et~al.} 2020,
  \mnras, 494, 4168

\bibitem[{{Dibi} {et~al.}(2014){Dibi}, {Markoff}, {Belmont}, {Malzac},
  {Barri{\`e}re}, \& {Tomsick}}]{Dibi2014}
{Dibi}, S., {Markoff}, S., {Belmont}, R., {et~al.} 2014, \mnras, 441, 1005

\bibitem[{Do {et~al.}(2009)Do, Ghez, Morris, Lu, Matthews, Yelda, \&
  Larkin}]{Do2009}
Do, T., Ghez, A.~M., Morris, M.~R., {et~al.} 2009, The Astrophysical Journal,
  703, 1323

\bibitem[{{Do} {et~al.}(2019){Do}, {Hees}, {Ghez}, {Martinez}, {Chu}, {Jia},
  {Sakai}, {Lu}, {Gautam}, {O'Neil}, {Becklin}, {Morris}, {Matthews},
  {Nishiyama}, {Campbell}, {Chappell}, {Chen}, {Ciurlo}, {Dehghanfar},
  {Gallego-Cano}, {Kerzendorf}, {Lyke}, {Naoz}, {Saida}, {Sch{\"o}del},
  {Takahashi}, {Takamori}, {Witzel}, \& {Wizinowich}}]{Do2019}
{Do}, T., {Hees}, A., {Ghez}, A., {et~al.} 2019, Science, 365, 664

\bibitem[{Dodds-Eden {et~al.}(2011)Dodds-Eden, Gillessen, Fritz, Eisenhauer,
  Trippe, Genzel, Ott, Bartko, Pfuhl, Bower, Goldwurm, Porquet, Trap, \&
  Yusef-Zadeh}]{Dodds-Eden2011}
Dodds-Eden, K., Gillessen, S., Fritz, T.~K., {et~al.} 2011, The Astrophysical
  Journal, 728, 37

\bibitem[{Dodds-Eden {et~al.}(2009)Dodds-Eden, Porquet, Trap, Quataert,
  Haubois, Gillessen, Grosso, Pantin, Falcke, Rouan, Genzel, Hasinger,
  Goldwurm, Yusef-Zadeh, Clenet, Trippe, Lagage, Bartko, Eisenhauer, Ott,
  Paumard, Perrin, Yuan, Fritz, \& Mascetti}]{Dodds-Eden2009}
Dodds-Eden, K., Porquet, D., Trap, G., {et~al.} 2009, The Astrophysical
  Journal, 698, 676

\bibitem[{Dodds-Eden {et~al.}(2010)Dodds-Eden, Sharma, Quataert, Genzel,
  Gillessen, Eisenhauer, \& Porquet}]{Dodds-Eden2010}
Dodds-Eden, K., Sharma, P., Quataert, E., {et~al.} 2010, The Astrophysical
  Journal, 725, 450

\bibitem[{{Eckart} {et~al.}(2004){Eckart}, {Baganoff}, {Morris}, {Bautz},
  {Brandt}, {Garmire}, {Genzel}, {Ott}, {Ricker}, {Straubmeier}, {Viehmann},
  {Sch{\"o}del}, {Bower}, \& {Goldston}}]{Eckart2004}
{Eckart}, A., {Baganoff}, F.~K., {Morris}, M., {et~al.} 2004, \aap, 427, 1

\bibitem[{{Eckart} {et~al.}(2009){Eckart}, {Baganoff}, {Morris}, {Kunneriath},
  {Zamaninasab}, {Witzel}, {Sch{\"o}del}, {Garc{\'\i}a-Mar{\'\i}n}, {Meyer},
  {Bower}, {Marrone}, {Bautz}, {Brandt}, {Garmire}, {Ricker}, {Straubmeier},
  {Roberts}, {Muzic}, {Mauerhan}, \& {Zensus}}]{Eckart2009}
{Eckart}, A., {Baganoff}, F.~K., {Morris}, M.~R., {et~al.} 2009, \aap, 500, 935

\bibitem[{{Eckart} {et~al.}(2012){Eckart}, {Garc{\'\i}a-Mar{\'\i}n}, {Vogel},
  {Teuben}, {Morris}, {Baganoff}, {Dexter}, {Sch{\"o}del}, {Witzel},
  {Valencia-S.}, {Karas}, {Kunneriath}, {Straubmeier}, {Moser}, {Sabha},
  {Buchholz}, {Zamaninasab}, {Mu{\v{z}}i{\'c}}, {Moultaka}, \&
  {Zensus}}]{Eckart2012}
{Eckart}, A., {Garc{\'\i}a-Mar{\'\i}n}, M., {Vogel}, S.~N., {et~al.} 2012,
  \aap, 537, A52

\bibitem[{{Eckart} {et~al.}(2008){Eckart}, {Sch{\"o}del},
  {Garc{\'\i}a-Mar{\'\i}n}, {Witzel}, {Weiss}, {Baganoff}, {Morris}, {Bertram},
  {Dov{\v{c}}iak}, {Duschl}, {Karas}, {K{\"o}nig}, {Krichbaum}, {Krips},
  {Kunneriath}, {Lu}, {Markoff}, {Mauerhan}, {Meyer}, {Moultaka},
  {Mu{\v{z}}i{\'c}}, {Najarro}, {Pott}, {Schuster}, {Sjouwerman},
  {Straubmeier}, {Thum}, {Vogel}, {Wiesemeyer}, {Zamaninasab}, \&
  {Zensus}}]{Eckart2008}
{Eckart}, A., {Sch{\"o}del}, R., {Garc{\'\i}a-Mar{\'\i}n}, M., {et~al.} 2008,
  \aap, 492, 337

\bibitem[{Eckart {et~al.}(2006)Eckart, Sch{\"{o}}del, Meyer, Trippe, Ott, \&
  Genzel}]{Eckart2006}
Eckart, A., Sch{\"{o}}del, R., Meyer, L., {et~al.} 2006, A\&A, 455, 1

\bibitem[{{Eisenhauer} {et~al.}(2005){Eisenhauer}, {Genzel}, {Alexander},
  {Abuter}, {Paumard}, {Ott}, {Gilbert}, {Gillessen}, {Horrobin}, {Trippe},
  {Bonnet}, {Dumas}, {Hubin}, {Kaufer}, {Kissler-Patig}, {Monnet},
  {Str{\"o}bele}, {Szeifert}, {Eckart}, {Sch{\"o}del}, \&
  {Zucker}}]{Eisenhauer2005}
{Eisenhauer}, F., {Genzel}, R., {Alexander}, T., {et~al.} 2005, \apj, 628, 246

\bibitem[{Falcke {et~al.}(1998)Falcke, Goss, Matsuo, Teuben, Zhao, \&
  Zylka}]{Falcke1998}
Falcke, H., Goss, W.~M., Matsuo, H., {et~al.} 1998, {The Simultaneous Spectrum
  of Sagittarius A* from 20 Centimeters to 1 Millimeter and the Nature of the
  Millimeter Excess}, Tech. Rep.~2

\bibitem[{{Fazio} {et~al.}(2018){Fazio}, {Ashby}, {Baganoff}, {Becklin},
  {Boyce}, {Carey}, {Gammie}, {Ghez}, {Glaccum}, {Gurwell}, {Haggard},
  {Herrero-Illana}, {Hora}, {Ingalls}, {Lowrance}, {Markoff}, {Marrone},
  {Morris}, {Narayan}, {Neilsen}, {Ponti}, {Smith}, {Willner}, \&
  {Witzel}}]{Fazio2018}
{Fazio}, G., {Ashby}, M., {Baganoff}, F., {et~al.} 2018, {The Vital Infrared to
  X-ray Link in the Sgr A* Accretion Flow}, Spitzer Proposal

\bibitem[{{Fazio} {et~al.}(2004){Fazio}, {Hora}, {Allen}, {Ashby}, {Barmby},
  {Deutsch}, {Huang}, {Kleiner}, {Marengo}, {Megeath}, {Melnick}, {Pahre},
  {Patten}, {Polizotti}, {Smith}, {Taylor}, {Wang}, {Willner}, {Hoffmann},
  {Pipher}, {Forrest}, {McMurty}, {McCreight}, {McKelvey}, {McMurray}, {Koch},
  {Moseley}, {Arendt}, {Mentzell}, {Marx}, {Losch}, {Mayman}, {Eichhorn},
  {Krebs}, {Jhabvala}, {Gezari}, {Fixsen}, {Flores}, {Shakoorzadeh}, {Jungo},
  {Hakun}, {Workman}, {Karpati}, {Kichak}, {Whitley}, {Mann}, {Tollestrup},
  {Eisenhardt}, {Stern}, {Gorjian}, {Bhattacharya}, {Carey}, {Nelson},
  {Glaccum}, {Lacy}, {Lowrance}, {Laine}, {Reach}, {Stauffer}, {Surace},
  {Wilson}, {Wright}, {Hoffman}, {Domingo}, \& {Cohen}}]{Fazio2004}
{Fazio}, G.~G., {Hora}, J.~L., {Allen}, L.~E., {et~al.} 2004, \apjs, 154, 10

\bibitem[{Fritz {et~al.}(2011)Fritz, Gillessen, Dodds-Eden, Lutz, Genzel, Raab,
  Ott, Pfuhl, Eisenhauer, \& Yusef-Zadeh}]{Fritz2011}
Fritz, T.~K., Gillessen, S., Dodds-Eden, K., {et~al.} 2011, The Astrophysical
  Journal, 737, 73

\bibitem[{{Fruscione} {et~al.}(2006){Fruscione}, {McDowell}, {Allen},
  {Brickhouse}, {Burke}, {Davis}, {Durham}, {Elvis}, {Galle}, {Harris},
  {Huenemoerder}, {Houck}, {Ishibashi}, {Karovska}, {Nicastro}, {Noble},
  {Nowak}, {Primini}, {Siemiginowska}, {Smith}, \& {Wise}}]{Fruscione2006}
{Fruscione}, A., {McDowell}, J.~C., {Allen}, G.~E., {et~al.} 2006, in Society
  of Photo-Optical Instrumentation Engineers (SPIE) Conference Series, Vol.
  6270, Society of Photo-Optical Instrumentation Engineers (SPIE) Conference
  Series, ed. D.~R. {Silva} \& R.~E. {Doxsey}, 62701V

\bibitem[{{Garmire} {et~al.}(2003){Garmire}, {Bautz}, {Ford}, {Nousek}, \&
  {Ricker}}]{Garmire2003}
{Garmire}, G.~P., {Bautz}, M.~W., {Ford}, P.~G., {Nousek}, J.~A., \& {Ricker},
  George~R., J. 2003, in Society of Photo-Optical Instrumentation Engineers
  (SPIE) Conference Series, Vol. 4851, X-Ray and Gamma-Ray Telescopes and
  Instruments for Astronomy., ed. J.~E. {Truemper} \& H.~D. {Tananbaum}, 28--44

\bibitem[{{Gehrels}(1986)}]{Gehrels1986}
{Gehrels}, N. 1986, \apj, 303, 336

\bibitem[{Genzel {et~al.}(2010)Genzel, Eisenhauer, \& Gillessen}]{Genzel2010}
Genzel, R., Eisenhauer, F., \& Gillessen, S. 2010, Reviews of Modern Physics,
  82

\bibitem[{Genzel {et~al.}(2003)Genzel, Sch{\"{o}}del, Ott, Eckart, Alexander,
  Lacombe, Rouan, \& Aschenbach}]{Genzel2003}
Genzel, R., Sch{\"{o}}del, R., Ott, T., {et~al.} 2003, Nature, 425, 934

\bibitem[{{Ghez} {et~al.}(2004){Ghez}, {Wright}, {Matthews}, {Thompson}, {Le
  Mignant}, {Tanner}, {Hornstein}, {Morris}, {Becklin}, \& {Soifer}}]{Ghez2004}
{Ghez}, A.~M., {Wright}, S.~A., {Matthews}, K., {et~al.} 2004, \apjl, 601, L159

\bibitem[{{Ghisellini}(2013)}]{Ghisellini2013}
{Ghisellini}, G. 2013, {Radiative Processes in High Energy Astrophysics}, Vol.
  873

\bibitem[{Gillessen {et~al.}(2006)Gillessen, Eisenhauer, Quataert, Genzel,
  Paumard, Trippe, Ott, Abuter, Eckart, Lagage, Lehnert, Tacconi, \&
  Martins}]{Gillessen2006}
Gillessen, S., Eisenhauer, F., Quataert, E., {et~al.} 2006, The Astrophysical
  Journal, 640, 163

\bibitem[{Gillessen {et~al.}(2017)Gillessen, Plewa, Eisenhauer, Sari, Waisberg,
  Habibi, Pfuhl, George, Dexter, von Fellenberg, Ott, \&
  Genzel}]{Gillessen2016}
Gillessen, S., Plewa, P.~M., Eisenhauer, F., {et~al.} 2017, The Astrophysical
  Journal, 837, 30

\bibitem[{{Gravity Collaboration} {et~al.}(2020{\natexlab{a}}){Gravity
  Collaboration}, Abuter, Amorim, Baub{\"{o}}ck, Berger, Bonnet, Brandner,
  Cardoso, Cl{\'{e}}net, {De Zeeuw}, Dallilar, Dexter, Eckart, Eisenhauer,
  {F{\"{o}}rster Schreiber}, Garcia, Gao, Gendron, Genzel, Gillessen, Habibi,
  Haubois, Henning, Hippler, Horrobin, Jim{\'{e}}nez-Rosales, Jochum, Jocou,
  Kaufer, Kervella, Lacour, Lapeyr{\`{e}}re, {Le Bouquin}, L{\'{e}}na, Nowak,
  Ott, Paumard, Perraut, Perrin, Pfuhl, Ponti, {Rodriguez Coira}, Shangguan,
  Scheithauer, Stadler, Straub, Straubmeier, Sturm, Tacconi, Vincent, {Von
  Fellenberg}, Waisberg, Widmann, Wieprecht, Wiezorrek, Woillez, Yazici, \&
  Zins}]{GRAVITYCollaboration2020flux}
{Gravity Collaboration}, Abuter, R., Amorim, A., {et~al.} 2020{\natexlab{a}},
  Astronomy and Astrophysics, 638

\bibitem[{{Gravity Collaboration} {et~al.}(2018){Gravity Collaboration},
  {Abuter}, {Amorim}, {Baub{\"o}ck}, {Berger}, {Bonnet}, {Brandner},
  {Cl{\'e}net}, {Coud{\'e} Du Foresto}, {de Zeeuw}, {Deen}, {Dexter}, {Duvert},
  {Eckart}, {Eisenhauer}, {F{\"o}rster Schreiber}, {Garcia}, {Gao}, {Gendron},
  {Genzel}, {Gillessen}, {Guajardo}, {Habibi}, {Haubois}, {Henning}, {Hippler},
  {Horrobin}, {Huber}, {Jim{\'e}nez-Rosales}, {Jocou}, {Kervella}, {Lacour},
  {Lapeyr{\`e}re}, {Lazareff}, {Le Bouquin}, {L{\'e}na}, {Lippa}, {Ott},
  {Panduro}, {Paumard}, {Perraut}, {Perrin}, {Pfuhl}, {Plewa}, {Rabien},
  {Rodr{\'\i}guez-Coira}, {Rousset}, {Sternberg}, {Straub}, {Straubmeier},
  {Sturm}, {Tacconi}, {Vincent}, {von Fellenberg}, {Waisberg}, {Widmann},
  {Wieprecht}, {Wiezorrek}, {Woillez}, \&
  {Yazici}}]{GravityCollaboration2018_orbital}
{Gravity Collaboration}, {Abuter}, R., {Amorim}, A., {et~al.} 2018, \aap, 618,
  L10

\bibitem[{{GRAVITY Collaboration} {et~al.}(2019){GRAVITY Collaboration},
  Abuter, Amorim, Baub{\"{o}}ck, Berger, Bonnet, Brandner, Cl{\'{e}}net,
  {Coud{\'{e}} du Foresto}, de~Zeeuw, Dexter, Duvert, Eckart, Eisenhauer,
  {F{\"{o}}rster Schreiber}, Garcia, Gao, Gendron, Genzel, Gerhard, Gillessen,
  Habibi, Haubois, Henning, Hippler, Horrobin, Jim{\'{e}}nez-Rosales, Jocou,
  Kervella, Lacour, Lapeyr{\`{e}}re, {Le Bouquin}, L{\'{e}}na, Ott, Paumard,
  Perraut, Perrin, Pfuhl, Rabien, {Rodriguez Coira}, Rousset, Scheithauer,
  Sternberg, Straub, Straubmeier, Sturm, Tacconi, Vincent, von Fellenberg,
  Waisberg, Widmann, Wieprecht, Wiezorrek, Woillez, \&
  Yazici}]{GRAVITYCollaboration2019}
{GRAVITY Collaboration}, Abuter, R., Amorim, A., {et~al.} 2019, \aap, 625, L10

\bibitem[{{Gravity Collaboration} {et~al.}(2021){Gravity Collaboration},
  {Abuter}, {Amorim}, {Baub{\"o}ck}, {Berger}, {Bonnet}, {Brandner},
  {Cl{\'e}net}, {Davies}, {de Zeeuw}, {Dexter}, {Dallilar}, {Drescher},
  {Eckart}, {Eisenhauer}, {F{\"o}rster Schreiber}, {Garcia}, {Gao}, {Gendron},
  {Genzel}, {Gillessen}, {Habibi}, {Haubois}, {Hei{\ss}el}, {Henning},
  {Hippler}, {Horrobin}, {Jim{\'e}nez-Rosales}, {Jochum}, {Jocou}, {Kaufer},
  {Kervella}, {Lacour}, {Lapeyr{\`e}re}, {Le Bouquin}, {L{\'e}na}, {Lutz},
  {Nowak}, {Ott}, {Paumard}, {Perraut}, {Perrin}, {Pfuhl}, {Rabien},
  {Rodr{\'\i}guez-Coira}, {Shangguan}, {Shimizu}, {Scheithauer}, {Stadler},
  {Straub}, {Straubmeier}, {Sturm}, {Tacconi}, {Vincent}, {von Fellenberg},
  {Waisberg}, {Widmann}, {Wieprecht}, {Wiezorrek}, {Woillez}, {Yazici},
  {Young}, \& {Zins}}]{GravityCollaboration2021}
{Gravity Collaboration}, {Abuter}, R., {Amorim}, A., {et~al.} 2021, \aap, 647,
  A59

\bibitem[{{Gravity Collaboration} {et~al.}(2020{\natexlab{b}}){Gravity
  Collaboration}, {Baub{\"o}ck}, {Dexter}, {Abuter}, {Amorim}, {Berger},
  {Bonnet}, {Brandner}, {Cl{\'e}net}, {Coud{\'e} Du Foresto}, {de Zeeuw},
  {Duvert}, {Eckart}, {Eisenhauer}, {F{\"o}rster Schreiber}, {Gao}, {Garcia},
  {Gendron}, {Genzel}, {Gerhard}, {Gillessen}, {Habibi}, {Haubois}, {Henning},
  {Hippler}, {Horrobin}, {Jim{\'e}nez-Rosales}, {Jocou}, {Kervella}, {Lacour},
  {Lapeyr{\`e}re}, {Le Bouquin}, {L{\'e}na}, {Ott}, {Paumard}, {Perraut},
  {Perrin}, {Pfuhl}, {Rabien}, {Rodriguez Coira}, {Rousset}, {Scheithauer},
  {Stadler}, {Sternberg}, {Straub}, {Straubmeier}, {Sturm}, {Tacconi},
  {Vincent}, {von Fellenberg}, {Waisberg}, {Widmann}, {Wieprecht}, {Wiezorrek},
  {Woillez}, \& {Yazici}}]{GravityCollaboration2020_orbital}
{Gravity Collaboration}, {Baub{\"o}ck}, M., {Dexter}, J., {et~al.}
  2020{\natexlab{b}}, \aap, 635, A143

\bibitem[{{Gravity Collaboration} {et~al.}(2020{\natexlab{c}}){Gravity
  Collaboration}, {Jim{\'e}nez-Rosales}, {Dexter}, {Widmann}, {Baub{\"o}ck},
  {Abuter}, {Amorim}, {Berger}, {Bonnet}, {Brandner}, {Cl{\'e}net}, {de Zeeuw},
  {Eckart}, {Eisenhauer}, {F{\"o}rster Schreiber}, {Garcia}, {Gao}, {Gendron},
  {Genzel}, {Gillessen}, {Habibi}, {Haubois}, {Hei{\ss}el}, {Henning},
  {Hippler}, {Horrobin}, {Jochum}, {Jocou}, {Kaufer}, {Kervella}, {Lacour},
  {Lapeyr{\`e}re}, {Le Bouquin}, {L{\'e}na}, {Nowak}, {Ott}, {Paumard},
  {Perraut}, {Perrin}, {Pfuhl}, {Rodr{\'\i}guez-Coira}, {Shangguan},
  {Scheithauer}, {Stadler}, {Straub}, {Straubmeier}, {Sturm}, {Tacconi},
  {Vincent}, {von Fellenberg}, {Waisberg}, {Wieprecht}, {Wiezorrek}, {Woillez},
  {Yazici}, \& {Zins}}]{GravityCollaboration2020_polariflares}
{Gravity Collaboration}, {Jim{\'e}nez-Rosales}, A., {Dexter}, J., {et~al.}
  2020{\natexlab{c}}, \aap, 643, A56

\bibitem[{{Harrison} {et~al.}(2013){Harrison}, {Craig}, {Christensen},
  {Hailey}, {Zhang}, {Boggs}, {Stern}, {Cook}, {Forster}, {Giommi},
  {Grefenstette}, {Kim}, {Kitaguchi}, {Koglin}, {Madsen}, {Mao}, {Miyasaka},
  {Mori}, {Perri}, {Pivovaroff}, {Puccetti}, {Rana}, {Westergaard}, {Willis},
  {Zoglauer}, {An}, {Bachetti}, {Barri{\`e}re}, {Bellm}, {Bhalerao},
  {Brejnholt}, {Fuerst}, {Liebe}, {Markwardt}, {Nynka}, {Vogel}, {Walton},
  {Wik}, {Alexander}, {Cominsky}, {Hornschemeier}, {Hornstrup}, {Kaspi},
  {Madejski}, {Matt}, {Molendi}, {Smith}, {Tomsick}, {Ajello}, {Ballantyne},
  {Balokovi{\'c}}, {Barret}, {Bauer}, {Blandford}, {Brandt}, {Brenneman},
  {Chiang}, {Chakrabarty}, {Chenevez}, {Comastri}, {Dufour}, {Elvis}, {Fabian},
  {Farrah}, {Fryer}, {Gotthelf}, {Grindlay}, {Helfand}, {Krivonos}, {Meier},
  {Miller}, {Natalucci}, {Ogle}, {Ofek}, {Ptak}, {Reynolds}, {Rigby},
  {Tagliaferri}, {Thorsett}, {Treister}, \& {Urry}}]{Harrison2013}
{Harrison}, F.~A., {Craig}, W.~W., {Christensen}, F.~E., {et~al.} 2013, \apj,
  770, 103

\bibitem[{Hora {et~al.}(2014)Hora, Witzel, Ashby, Becklin, Carey, Fazio, Ghez,
  Ingalls, Meyer, Morris, Smith, \& Willner}]{Hora2014}
Hora, J.~L., Witzel, G., Ashby, M.~L., {et~al.} 2014, Astrophysical Journal,
  793, 120

\bibitem[{{Hornstein} {et~al.}(2007){Hornstein}, {Matthews}, {Ghez}, {Lu},
  {Morris}, {Becklin}, {Rafelski}, \& {Baganoff}}]{Hornstein2007}
{Hornstein}, S.~D., {Matthews}, K., {Ghez}, A.~M., {et~al.} 2007, \apj, 667,
  900

\bibitem[{{Ingalls} {et~al.}(2012){Ingalls}, {Krick}, {Carey}, {Laine},
  {Surace}, {Glaccum}, {Grillmair}, \& {Lowrance}}]{Ingalls2012}
{Ingalls}, J.~G., {Krick}, J.~E., {Carey}, S.~J., {et~al.} 2012, in Society of
  Photo-Optical Instrumentation Engineers (SPIE) Conference Series, Vol. 8442,
  Space Telescopes and Instrumentation 2012: Optical, Infrared, and Millimeter
  Wave, ed. M.~C. {Clampin}, G.~G. {Fazio}, H.~A. {MacEwen}, \& J.~{Oschmann},
  Jacobus~M., 84421Y

\bibitem[{{Issaoun} {et~al.}(2019){Issaoun}, {Johnson}, {Blackburn},
  {Brinkerink}, {Mo{\'s}cibrodzka}, {Chael}, {Goddi}, {Mart{\'\i}-Vidal},
  {Wagner}, {Doeleman}, {Falcke}, {Krichbaum}, {Akiyama}, {Bach}, {Bouman},
  {Bower}, {Broderick}, {Cho}, {Crew}, {Dexter}, {Fish}, {Gold}, {G{\'o}mez},
  {Hada}, {Hern{\'a}ndez-G{\'o}mez}, {Jan{\ss}en}, {Kino}, {Kramer}, {Loinard},
  {Lu}, {Markoff}, {Marrone}, {Matthews}, {Moran}, {M{\"u}ller}, {Roelofs},
  {Ros}, {Rottmann}, {Sanchez}, {Tilanus}, {de Vicente}, {Wielgus}, {Zensus},
  \& {Zhao}}]{Issaoun2019}
{Issaoun}, S., {Johnson}, M.~D., {Blackburn}, L., {et~al.} 2019

\bibitem[{{Jin} {et~al.}(2017){Jin}, {Ponti}, {Haberl}, \& {Smith}}]{Jin2017}
{Jin}, C., {Ponti}, G., {Haberl}, F., \& {Smith}, R. 2017, \mnras, 468, 2532

\bibitem[{{Jin} {et~al.}(2018){Jin}, {Ponti}, {Haberl}, {Smith}, \&
  {Valencic}}]{Jin2018}
{Jin}, C., {Ponti}, G., {Haberl}, F., {Smith}, R., \& {Valencic}, L. 2018,
  \mnras, 477, 3480

\bibitem[{{Kardashev}(1962)}]{Kardashev1962}
{Kardashev}, N.~S. 1962, \sovast, 6, 317

\bibitem[{{Li} {et~al.}(2017){Li}, {Yuan}, \& {Wang}}]{Li2017}
{Li}, Y.-P., {Yuan}, F., \& {Wang}, Q.~D. 2017, \mnras, 468, 2552

\bibitem[{Liu {et~al.}(2016)Liu, Wright, Zhao, Brinkerink, {T. P. Ho}, Mills,
  Mart{\'{i}}n, Falcke, Matsushita, \& Mart{\'{i}}-Vidal}]{Liu2016}
Liu, H.~B., Wright, M. C.~H., Zhao, J.-H., {et~al.} 2016, \aap, 593, A107

\bibitem[{{Loeb} \& {Waxman}(2007)}]{LoebWaxman2007}
{Loeb}, A. \& {Waxman}, E. 2007, \jcap, 2007, 011

\bibitem[{{Malyshev} {et~al.}(2015){Malyshev}, {Chernyakova}, {Neronov}, \&
  {Walter}}]{Malyshev2015}
{Malyshev}, D., {Chernyakova}, M., {Neronov}, A., \& {Walter}, R. 2015, \aap,
  582, A11

\bibitem[{{Markoff} {et~al.}(2001){Markoff}, {Falcke}, {Yuan}, \&
  {Biermann}}]{Markoff2001}
{Markoff}, S., {Falcke}, H., {Yuan}, F., \& {Biermann}, P.~L. 2001, \aap, 379,
  L13

\bibitem[{{Marrone} {et~al.}(2008){Marrone}, {Baganoff}, {Morris}, {Moran},
  {Ghez}, {Hornstein}, {Dowell}, {Mu{\~n}oz}, {Bautz}, {Ricker}, {Brandt},
  {Garmire}, {Lu}, {Matthews}, {Zhao}, {Rao}, \& {Bower}}]{Marrone2008}
{Marrone}, D.~P., {Baganoff}, F.~K., {Morris}, M.~R., {et~al.} 2008, \apj, 682,
  373

\bibitem[{Meyer {et~al.}(2009)Meyer, Do, Ghez, Morris, Yelda, Sch{\"{o}}del, \&
  Eckart}]{Meyer2009}
Meyer, L., Do, T., Ghez, A., {et~al.} 2009, The Astrophysical Journal, 694, 87

\bibitem[{Mo{\'{s}}cibrodzka \& Falcke(2013)}]{Moscibrodzka2013}
Mo{\'{s}}cibrodzka, M. \& Falcke, H. 2013, \aap, 559, L3

\bibitem[{{Mossoux} {et~al.}(2016){Mossoux}, {Grosso}, {Bushouse}, {Eckart},
  {Yusef-Zadeh}, {Plambeck}, {Peissker}, {Valencia-S.}, {Porquet}, {Cotton}, \&
  {Roberts}}]{2016A&A...589A.116M}
{Mossoux}, E., {Grosso}, N., {Bushouse}, H., {et~al.} 2016, \aap, 589, A116

\bibitem[{{Neilsen} {et~al.}(2013){Neilsen}, {Nowak}, {Gammie}, {Dexter},
  {Markoff}, {Haggard}, {Nayakshin}, {Wang}, {Grosso}, {Porquet}, {Tomsick},
  {Degenaar}, {Fragile}, {Houck}, {Wijnands}, {Miller}, \&
  {Baganoff}}]{Neilsen2013}
{Neilsen}, J., {Nowak}, M.~A., {Gammie}, C., {et~al.} 2013, \apj, 774, 42

\bibitem[{{Nowak} {et~al.}(2012){Nowak}, {Neilsen}, {Markoff}, {Baganoff},
  {Porquet}, {Grosso}, {Levin}, {Houck}, {Eckart}, {Falcke}, {Ji}, {Miller}, \&
  {Wang}}]{Nowak2012}
{Nowak}, M.~A., {Neilsen}, J., {Markoff}, S.~B., {et~al.} 2012, \apj, 759, 95

\bibitem[{{Pacholczyk}(1970)}]{Pacholczyk1970}
{Pacholczyk}, A.~G. 1970, {Radio astrophysics. Nonthermal processes in galactic
  and extragalactic sources}

\bibitem[{{Pandya} {et~al.}(2016){Pandya}, {Zhang}, {Chandra}, \&
  {Gammie}}]{Pandya2016}
{Pandya}, A., {Zhang}, Z., {Chandra}, M., \& {Gammie}, C.~F. 2016, \apj, 822,
  34

\bibitem[{{Ponti} {et~al.}(2015){Ponti}, {De Marco}, {Morris}, {Merloni},
  {Mu{\~n}oz-Darias}, {Clavel}, {Haggard}, {Zhang}, {Nandra}, {Gillessen},
  {Mori}, {Neilsen}, {Rea}, {Degenaar}, {Terrier}, \& {Goldwurm}}]{Ponti2015}
{Ponti}, G., {De Marco}, B., {Morris}, M.~R., {et~al.} 2015, \mnras, 454, 1525

\bibitem[{Ponti {et~al.}(2017)Ponti, George, Scaringi, Zhang, Jin, Dexter,
  Terrier, Clavel, Degenaar, Eisenhauer, Genzel, Gillessen, Goldwurm, Habibi,
  Haggard, Hailey, Harrison, Merloni, Mori, Nandra, Ott, Pfuhl, Plewa, \&
  Waisberg}]{Ponti2017}
Ponti, G., George, E., Scaringi, S., {et~al.} 2017, Monthly Notices of the
  Royal Astronomical Society, 468, 2447

\bibitem[{{Porquet} {et~al.}(2008){Porquet}, {Grosso}, {Predehl}, {Hasinger},
  {Yusef-Zadeh}, {Aschenbach}, {Trap}, {Melia}, {Warwick}, {Goldwurm},
  {B{\'e}langer}, {Tanaka}, {Genzel}, {Dodds-Eden}, {Sakano}, \&
  {Ferrando}}]{Porquet2008}
{Porquet}, D., {Grosso}, N., {Predehl}, P., {et~al.} 2008, \aap, 488, 549

\bibitem[{{Porquet} {et~al.}(2003){Porquet}, {Predehl}, {Aschenbach}, {Grosso},
  {Goldwurm}, {Goldoni}, {Warwick}, \& {Decourchelle}}]{Porquet2003}
{Porquet}, D., {Predehl}, P., {Aschenbach}, B., {et~al.} 2003, \aap, 407, L17

\bibitem[{{Quataert}(2002)}]{Quataert2002}
{Quataert}, E. 2002, \apj, 575, 855

\bibitem[{{Ripperda} {et~al.}(2020){Ripperda}, {Bacchini}, \&
  {Philippov}}]{Ripperda2020}
{Ripperda}, B., {Bacchini}, F., \& {Philippov}, A.~A. 2020, \apj, 900, 100

\bibitem[{{Sironi} \& {Beloborodov}(2020)}]{Sironi2020}
{Sironi}, L. \& {Beloborodov}, A.~M. 2020, \apj, 899, 52

\bibitem[{Stone {et~al.}(2016)Stone, Marrone, Dowell, Schulz, Heinke, \&
  Yusef-Zadeh}]{Stone2016}
Stone, J.~M., Marrone, D.~P., Dowell, C.~D., {et~al.} 2016, The Astrophysical
  Journal, 825, 32

\bibitem[{{Trap} {et~al.}(2011){Trap}, {Goldwurm}, {Dodds-Eden}, {Weiss},
  {Terrier}, {Ponti}, {Gillessen}, {Genzel}, {Ferrando}, {B{\'e}langer},
  {Cl{\'e}net}, {Rouan}, {Predehl}, {Capelli}, {Melia}, \&
  {Yusef-Zadeh}}]{Trap2011}
{Trap}, G., {Goldwurm}, A., {Dodds-Eden}, K., {et~al.} 2011, \aap, 528, A140

\bibitem[{{Verner} {et~al.}(1996){Verner}, {Ferland}, {Korista}, \&
  {Yakovlev}}]{Verner1996}
{Verner}, D.~A., {Ferland}, G.~J., {Korista}, K.~T., \& {Yakovlev}, D.~G. 1996,
  \apj, 465, 487

\bibitem[{von Fellenberg {et~al.}(2018)von Fellenberg, Gillessen,
  Graci{\'{a}}-Carpio, Fritz, Dexter, Baub{\"{o}}ck, Ponti, Gao, Habibi, Plewa,
  Pfuhl, Jimenez-Rosales, Waisberg, Widmann, Ott, Eisenhauer, \&
  Genzel}]{VonFellenberg2018}
von Fellenberg, S.~D., Gillessen, S., Graci{\'{a}}-Carpio, J., {et~al.} 2018,
  The Astrophysical Journal, 862, 129

\bibitem[{{Weisskopf} {et~al.}(2000){Weisskopf}, {Tananbaum}, {Van Speybroeck},
  \& {O'Dell}}]{Weisskopf2000}
{Weisskopf}, M.~C., {Tananbaum}, H.~D., {Van Speybroeck}, L.~P., \& {O'Dell},
  S.~L. 2000, in Society of Photo-Optical Instrumentation Engineers (SPIE)
  Conference Series, Vol. 4012, X-Ray Optics, Instruments, and Missions III,
  ed. J.~E. {Truemper} \& B.~{Aschenbach}, 2--16

\bibitem[{{Werner} \& {Uzdensky}(2021)}]{Werner2021}
{Werner}, G.~R. \& {Uzdensky}, D.~A. 2021, arXiv e-prints, arXiv:2106.02790

\bibitem[{{Werner} {et~al.}(2004){Werner}, {Roellig}, {Low}, {Rieke}, {Rieke},
  {Hoffmann}, {Young}, {Houck}, {Brandl}, {Fazio}, {Hora}, {Gehrz}, {Helou},
  {Soifer}, {Stauffer}, {Keene}, {Eisenhardt}, {Gallagher}, {Gautier}, {Irace},
  {Lawrence}, {Simmons}, {Van Cleve}, {Jura}, {Wright}, \&
  {Cruikshank}}]{Werner2004}
{Werner}, M.~W., {Roellig}, T.~L., {Low}, F.~J., {et~al.} 2004, \apjs, 154, 1

\bibitem[{{Wilms} {et~al.}(2000{\natexlab{a}}){Wilms}, {Allen}, \&
  {McCray}}]{WilmsAllenMcCray2000}
{Wilms}, J., {Allen}, A., \& {McCray}, R. 2000{\natexlab{a}}, \apj, 542, 914

\bibitem[{{Wilms} {et~al.}(2000{\natexlab{b}}){Wilms}, {Allen}, \&
  {McCray}}]{Wilms2000}
{Wilms}, J., {Allen}, A., \& {McCray}, R. 2000{\natexlab{b}}, \apj, 542, 914

\bibitem[{Witzel {et~al.}(2018)Witzel, Martinez, Hora, Willner, Morris, Gammie,
  Becklin, Ashby, Baganoff, Carey, Do, Fazio, Ghez, Glaccum, Haggard,
  Herrero-Illana, Ingalls, Narayan, \& Smith}]{Witzel2018}
Witzel, G., Martinez, G., Hora, J., {et~al.} 2018, The Astrophysical Journal,
  863, 15

\bibitem[{{Wong} {et~al.}(2020){Wong}, {Zhdankin}, {Uzdensky}, {Werner}, \&
  {Begelman}}]{Wong2020}
{Wong}, K., {Zhdankin}, V., {Uzdensky}, D.~A., {Werner}, G.~R., \& {Begelman},
  M.~C. 2020, \apjl, 893, L7

\bibitem[{{Xu} {et~al.}(2006){Xu}, {Narayan}, {Quataert}, {Yuan}, \&
  {Baganoff}}]{Xu2006}
{Xu}, Y.-D., {Narayan}, R., {Quataert}, E., {Yuan}, F., \& {Baganoff}, F.~K.
  2006, \apj, 640, 319

\bibitem[{{Yoon} {et~al.}(2020){Yoon}, {Chatterjee}, {Markoff}, {van
  Eijnatten}, {Younsi}, {Liska}, \& {Tchekhovskoy}}]{2020MNRAS.499.3178Y}
{Yoon}, D., {Chatterjee}, K., {Markoff}, S.~B., {et~al.} 2020, \mnras, 499,
  3178

\bibitem[{Yuan {et~al.}(2003)Yuan, Quataert, \& Narayan}]{Yuan2003}
Yuan, F., Quataert, E., \& Narayan, R. 2003, The Astrophysical Journal, 598,
  301

\bibitem[{{Yusef-Zadeh} {et~al.}(2009){Yusef-Zadeh}, {Bushouse}, {Wardle},
  {Heinke}, {Roberts}, {Dowell}, {Brunthaler}, {Reid}, {Martin}, {Marrone},
  {Porquet}, {Grosso}, {Dodds-Eden}, {Bower}, {Wiesemeyer}, {Miyazaki}, {Pal},
  {Gillessen}, {Goldwurm}, {Trap}, \& {Maness}}]{YusefZadeh2009}
{Yusef-Zadeh}, F., {Bushouse}, H., {Wardle}, M., {et~al.} 2009, \apj, 706, 348

\bibitem[{{Yusef-Zadeh} {et~al.}(2006){Yusef-Zadeh}, {Roberts}, {Wardle},
  {Heinke}, \& {Bower}}]{YusefZadeh2006}
{Yusef-Zadeh}, F., {Roberts}, D., {Wardle}, M., {Heinke}, C.~O., \& {Bower},
  G.~C. 2006, \apj, 650, 189

\bibitem[{{Yusef-Zadeh} {et~al.}(2008){Yusef-Zadeh}, {Wardle}, {Heinke},
  {Dowell}, {Roberts}, {Baganoff}, \& {Cotton}}]{YusefZadeh2008}
{Yusef-Zadeh}, F., {Wardle}, M., {Heinke}, C., {et~al.} 2008, \apj, 682, 361

\end{thebibliography}

\bibliographystyle{aa}

\end{document}